\documentclass[preprint]{aastex}
\usepackage{graphicx}
\usepackage{amsmath}

\shorttitle{Red Optical Variability of T dwarfs}
\shortauthors{Heinze et al.}

\begin{document}

\title{Weather on Other Worlds. III. A Survey for T Dwarfs with High Amplitude Optical Variability}

\author{Aren N. Heinze\altaffilmark{1}, Stanimir Metchev\altaffilmark{2,1}, Kendra Kellogg\altaffilmark{2}}

\altaffiltext{1}{Department of Physics and Astronomy, Stony Brook University, Stony Brook, NY 11794-3800, USA; aren.heinze@stonybrook.edu}
\altaffiltext{2}{Department of Physics and Astronomy, The University of Western Ontario, 1151 Richmond St, London, ON N6A 3K7, Canada; smetchev@uwo.ca}

\begin{abstract}
We have monitored twelve T dwarfs with the Kitt Peak 2.1m telescope using an f814w filter (0.7-0.95 $\mu$m) to 
place in context the remarkable 10-20\% variability exhibited by the nearby T dwarf
Luhman 16B in this wavelength regime.  The motivation was the poorly known red optical behavior of T dwarfs, which have been
monitored almost exclusively at infrared wavelengths, where variability
amplitudes greater than 10\% have been found to be very rare.  
We detect highly significant variability in two T dwarfs.
The T2.5 dwarf 2MASS 13243559+6358284
shows consistent $\sim$17\% variability on two consecutive nights.  
The T2 dwarf 2MASS J16291840+0335371 exhibits $\sim$10\% variability that may evolve from night to night, similarly
to Luhman 16B.  Both objects were previously known to be variable in the infrared, 
but with considerably lower amplitudes.  
We also find evidence for variability in the T6 dwarf J162414.37+002915.6, but since
it has lower significance, we conservatively refrain from claiming this object as a variable.
We explore and rule out various telluric effects, demonstrating
that the variations we detect are astrophysically real.
We suggest that high-amplitude photometric
variability for T dwarfs is likely more common in the red optical
than at longer wavelengths. The two new members of the growing class of high-amplitude
variable T dwarfs offer excellent prospects for further study of cloud
structures and their evolution.
\end{abstract}


\section{Introduction}

The nearest T dwarf, the T1.5 secondary component \citep{Kniazev2013} in the L8+T1.5 binary Luhman 16 
(WISE J1049; Luhman 2013), is one of the most highly variable brown dwarfs.   
\citet{Gillon2013} found Luhman 16B to vary with a period of $4.87 \pm 0.01$ hr and an 
amplitude exceeding
20\% in the optical on some nights. Among periodically varying brown dwarfs, this amplitude is
exceeded only by the 26\% $J$-band variability detected by \citet{2M2139} in 2MASS J21392676+0220226,
which was the only one of over 150 brown dwarfs monitored in surveys by \citet{Koen2013a},
\citet{Radigan2014}, and ourselves \citep{Metchev2014} that exhibited periodic variations with
an amplitude greater than 10\%.  Even taking into account
its inconstant amplitude, Luhman 16B appears as an outlier in this context,
with an amplitude at roughly the 99th percentile for brown dwarfs. 

However, none of
these previous surveys has focused on T dwarfs in the red optical, where Luhman 16B has its
largest measured amplitude.  \citet{Radigan2014} and \citet{Metchev2014} targeted
both L and T dwarfs, but observed exclusively in the infrared (the $J$ band and the IRAC
3.6 and 4.5 $\mu$m channels, respectively). \citet{Koen2013a} used the red optical
$I$ and $R$ bands, but observed almost exclusively L and M dwarfs, including only three
T dwarfs in his sample. Large-amplitude variability is rare for both L and T dwarfs
in the infrared, and for L dwarfs in the red optical --- but the red optical behavior
of T dwarfs has not been systematically explored.  Without
such an exploration, we lack context for understanding Luhman 16B's photometric behavior, and
we cannot say how representative the nearest T dwarf is of its spectral type.
We report herein the results of the first survey aimed at providing such context.

\section{Survey Design} \label{sec:survey}

We aim to determine the occurrence rate, among T dwarfs, of objects with periodic red-optical
variability $\ge10$\%. In the context of our current
investigation we will also refer to this as `high amplitude variability'.
Photometry of sufficient precision to meet our goals is possible
only for a relatively small number of bright T dwarfs, so we selected
our targets primarily based on flux.  Our only other requirement was 
that our targets must spend at least six hours
above airmass 2.0 as seen from Kitt Peak in the first half of the year.
The time-of-year constraint was needed because all our observations had to be conducted
during the 2014A observing semester, after which the Kitt Peak 2.1m was divested from NOAO.

Many T dwarfs lack published $I$-band photometry, but their $I$-band magnitudes
can be estimated to within about $\pm 0.4$ mag from $J$-band photometry and typical
$I-J$ colors, or from SDSS $i$ and $z$ magnitudes, where available.
We found that the twelve brightest T dwarfs listed in DwarfArchives.org and observable from 
Kitt Peak in January--June had estimated $I$ band magnitudes of $\sim 20$ or brighter.  Our own previous
2.1m observations had shown that $I \sim 20$ mag objects could be monitored with sufficient
precision for our purposes; therefore these twelve objects became our survey sample.  They
are listed in Table \ref{tab:targs}, together with the shorthand names we will use
to refer to them herein.  Table \ref{tab:targs} also gives the first non-SDSS optical
magnitudes for most of our targets, based not on our selection estimates but on our 
own calibrated photometry. They give
context for our photometric precision (Section \ref{sec:phot} and Table \ref{tab:data}),
and may be useful in planning future observations.

Our sample is unbiased with respect to known variability or inclusion
in previous surveys, and includes seven objects previously monitored in the
infrared: SDSS 1254 (both Radigan et al. 2014 and Metchev et al. 2014); SDSS 1207,
2M 1231, SDSS 1624, and 2M 1629 \citep{Radigan2014}; and 2M 1324 and SDSS 1520 \citep{Metchev2014}.  
Of these, 2M 1324 and 2M 1629 were reported variable, in the IRAC 3--5 $\mu$m bands
and the $J$ band respectively.  None
of our targets was previously monitored in the red optical.

\begin{deluxetable}{lllcccccc}
\rotate
\tabletypesize{\footnotesize}
\tablewidth{0pt}
\tablecaption{Target Objects \label{tab:targs}}
\tablehead{\colhead{Object}  & \colhead{Short} & \colhead{Spec.} & \multicolumn{6}{c}{Magnitudes} \\
\colhead{Designation} & \colhead{Name} & \colhead{Type} & \colhead{$J$} & \colhead{$i$\tablenotemark{a}} & \colhead{$z$\tablenotemark{a}} & \colhead{f814w\tablenotemark{b}} & \colhead{$I$\tablenotemark{b}} & \colhead{$R$\tablenotemark{b}} }
\startdata
WISEPA J081958.05-033529.0 & WISE 0819 & T4\tablenotemark{c} & $14.99 \pm 0.04$\tablenotemark{c} & \nodata & \nodata & $19.46 \pm 0.02$ & $19.73 \pm 0.04$ &  $23.3 \pm 0.2$ \\
2MASSI J0937347+293142 & 2M 0937 & T6p\tablenotemark{e} & $14.65 \pm 0.04$\tablenotemark{d} & $22.00 \pm 0.28$ & $17.50 \pm 0.02$ & $19.53 \pm 0.02$ & $19.72 \pm 0.02$ &  $23.5 \pm 0.3$ \\
2MASS J11061197+2754225 & 2M 1106 & T2.5\tablenotemark{f} & $14.82 \pm 0.04$\tablenotemark{d} & $21.24 \pm 0.10$ & $17.70 \pm 0.02$ & $18.80 \pm 0.02$ & $18.95 \pm 0.02$ &  $22.2 \pm 0.1$ \\
SDSS J120747.17+024424.8 & SDSS 1207 & T0\tablenotemark{e} & $15.58 \pm 0.07$\tablenotemark{d} & $21.44 \pm 0.11$ & $18.39 \pm 0.04$ & $19.27 \pm 0.03$ & $19.39 \pm 0.03$ &  $22.2 \pm 0.1$ \\
2MASS J12314753+0847331 & 2M 1231 & T5.5\tablenotemark{e} & $15.57 \pm 0.07$\tablenotemark{d} & $22.76 \pm 0.29$ & $18.91 \pm 0.04$ & $20.11 \pm 0.03$ & $20.31 \pm 0.03$ &  $24.1 \pm 0.5$ \\
SDSSp J125453.90-012247.4 & SDSS 1254 & T2\tablenotemark{e} & $14.89 \pm 0.04$\tablenotemark{d} & $22.18 \pm 0.27$ & $18.02 \pm 0.03$ &  $19.11 \pm 0.02$ & $19.28 \pm 0.02$ &  \nodata \\
2MASS J13243553+6358281 & 2M 1324 & T2.5\tablenotemark{f,}\tablenotemark{g} & $15.60 \pm 0.07$\tablenotemark{d} & $22.74 \pm 0.27$ & $18.74 \pm 0.04$ & $19.87 \pm 0.05$ & \nodata &  \nodata \\
2MASS J15031961+2525196 & 2M 1503 & T5\tablenotemark{e} & $13.94 \pm 0.02$\tablenotemark{d} & $22.02 \pm 0.16$ & $17.30 \pm 0.01$ & $18.39 \pm 0.05$ & \nodata &  \nodata \\
WISEPC J150649.97+702736.0 & WISE 1506 & T6\tablenotemark{c} & $14.33 \pm 0.10$\tablenotemark{c} & \nodata & \nodata &  $18.55 \pm 0.05$ & \nodata &  \nodata \\
SDSS J152039.82+354619.8 & SDSS 1520 & T0\tablenotemark{h} & $15.54 \pm 0.06$\tablenotemark{d} & $21.88 \pm 0.12$ & $18.35 \pm 0.02$ & $19.22 \pm 0.07$ & \nodata &  $22.4 \pm 0.1$ \\
SDSSp J162414.37+002915.6 & SDSS 1624 & T6\tablenotemark{e} & $15.49 \pm 0.05$\tablenotemark{d} & $22.82 \pm 0.27$ & $19.07 \pm 0.04$ & $20.30 \pm 0.04$ & \nodata &  $>23.6$\tablenotemark{j} \\
2MASS J16291840+0335371 & 2M 1629 & T2\tablenotemark{i} & $15.29 \pm 0.04$\tablenotemark{d} & \nodata & \nodata & $19.59 \pm 0.04$ & \nodata &  $23.2 \pm 0.2$ \\
\enddata
\tablenotetext{a}{From SDSS \citep{SDSS}.}
\tablenotetext{b}{From this work.  The $R$ and $I$ magnitudes are calibrated
from standard stars in \citet{Landolt1992}. The f814w magnitudes
are effectively on a unique photometric system defined by the $I$-band magnitudes of the
Landolt standards, but they can be used to predict $I$-band fluxes with
greater accuracy than can be achieved from $J$-band magnitudes or SDSS $i$ and $z$ photometry, and
thus they may be useful for planning future observations.}
\tablenotetext{c}{From \citet{Kirkpatrick2011}.}
\tablenotetext{d}{From 2MASS \citep{2MASS}.}
\tablenotetext{e}{\citet{Burgasser2006}}
\tablenotetext{f}{\citet{Looper2007}}
\tablenotetext{g}{\citet{Metchev2008}}
\tablenotetext{h}{\citet{Chiu2006}}
\tablenotetext{i}{\citet{Deacon2011}}
\tablenotetext{j}{Target not detected in $R$; quoted limit is 2$\sigma$.}
\end{deluxetable}

\section{Observations and Image Processing} \label{sec:Data}

\subsection{Observations} \label{sec:Obs}
Our observations were designed to detect variations with a 10\% or larger
amplitude that might not be consistent from night to night, at any period in the typical range
for brown dwarf rotations.  Such periods range from two to twelve hours
\citep{Zapatero2006,Reiners2008a,Konopacky2012}, though a few both shorter and
longer can be found \citep{Metchev2014}. To achieve this, we spent
at least two nights on each of our survey targets, acquiring the longest possible
temporal baseline on each night. For some objects we were able to obtain
three or four nights' monitoring to enhance the confidence of variability
detections and better constrain long periods(see Table \ref{tab:data}).

All of our data were obtained during four observing runs in 2014, which comprised a total of 22 nights: 
four in February, eight in March, and five each in May and June. We
used the Kitt Peak 2.1m telescope's STA3 facility CCD imager, which has a plate scale of 0.305 pixels/arcsec
and a 6x10 arcminute field of view.
For our photometric monitoring we used the KPNO 1560 filter, which
is a Hubble Space Telescope f814w filter similar to the I band, but wider, delivering
better signal to noise on brown dwarfs.  On twelve nights we observed one primary
target continuously except for brief visits to calibration fields or targets of
opportunity and (on some nights) intervals of cloud.  On the remaining ten nights we alternated
between two brown dwarf targets, acquiring temporal baselines spanning most of
the night on each.  
Table \ref{tab:data} details our photometric monitoring sequences for each object.

Some of our targets were not easy to identify in the f814w images
due to their faintness and rapid proper motion.  We obtained $R$-band images
of these, which allowed us to confidently identify the T dwarfs by their very
red $R - \mathrm{f814w}$ colors, and also to measure the $R$-band magnitudes given
in Table \ref{tab:targs}.  We used a consistent exposure time of 200 seconds for all
images of our targets.

Our photometric monitoring observations employed a spatially confined non-redundant
dither pattern.  It has the properties that no two images on a given night
are taken at the same telescope pointing; all pointings cluster within
a 40 arcsecond radius of a nominal center; and the direction of the offset
from the nominal center is reversed every five images or more often. This technique
has several advantages: stars mostly disappear on a clipped median-stack of
background-normalized images, resulting in a clean sky/fringe frame; the spatial
confinement of the dither pattern ensures that a large-scale residual flatfield gradient
cannot significantly affect the photometry; and the alternation
between different offset directions from the nominal center is more rapid
that any expected astrophysical signal.  This last property ensures that
flatfield gradients or other systematic effects with dependence on detector
position are not covariant with any true astrophysical variability.
An alternative method, `staring', has
been extensively used for precision photometric monitoring \citep{Deming12,Radigan2014},
but this is appropriate mainly when small-scale flatfielding residuals are an
important source of photometric error.  Such residuals are usually important in infrared
arrays, but in astronomical CCDs they are much smaller and become significant
only for objects far brighter than our targets. For observations such as ours,
dithering allows better sky subtraction than staring, and this drives our choice.

\begin{deluxetable}{ccccccccc}
\tablecaption{Photometric Monitoring Data Acquired \label{tab:data}}
\tabletypesize{\footnotesize}
\tablehead{  & &  &  & \colhead{Sky}  & \colhead{Apert./} & \colhead{Sky Sub.} & & \colhead{Offset}\\
\colhead{Target} & \colhead{UT Date} & \colhead{Images\tablenotemark{a}}& \colhead{Objects\tablenotemark{b}} & \colhead{Conditions} & \colhead{Seeing\tablenotemark{c}} & \colhead{Method} & \colhead{RMS\tablenotemark{d}} & \colhead{of Mean\tablenotemark{e}}}
\startdata
WISE 0819 & 2014-02-18 & 43 & 77 & clear & 1.83/1.63 & none\tablenotemark{g} & 2.9\% & -1.9\% \\
WISE 0819 & 2014-02-20 & 47 & 77 & clear & 2.14/2.15 & single-frame & 3.9\% & +3.2\% \\
WISE 0819 & 2014-02-21 & 41 & 77 & some clouds & 1.53/1.55 & multi-frame & 2.2\% & +0.6\% \\
2M 0937 & 2014-02-18 & 55 & 28 & clear & 2.14/1.75 & multi-frame\tablenotemark{g} & 4.6\% & -1.1\% \\
2M 0937 & 2014-02-19 & 92 & 28 & some clouds & 2.14/1.72 & single-frame & 3.6\% & -0.7\% \\
2M 0937 & 2014-02-20 & 67 & 28 & clear & 2.14/2.02 & multi-frame & 2.8\% & +2.3\% \\
2M 0937 & 2014-02-21 & 48 & 28 & some clouds & 1.83/1.51 & single-frame & 2.8\% & +0.2\% \\
2M 1106 & 2014-03-07 & 100 & 17 & some clouds & 2.14/1.64 & multi-frame\tablenotemark{g} & 1.2\% & +0.1\% \\
2M 1106 & 2014-03-08 & 38 & 17  & some clouds & 1.83/1.58 & multi-frame\tablenotemark{g} & 1.4\% & +0.2\% \\
SDSS 1207 & 2014-03-08 & 55 & 19 & some clouds & 1.83/1.80 & multi-frame\tablenotemark{g} & 2.1\% & -0.9\% \\
SDSS 1207 & 2014-03-10 & 47 & 19 & clear & 2.14/1.79 & multi-frame & 1.9\% & +0.8\% \\
2M 1231 & 2014-03-10 & 45 & 22 & clear & 1.83/1.81 & multi-frame & 4.6\% & +0.3\% \\
2M 1231 & 2014-03-11 & 77 & 22 & clear & 2.44/2.08 & multi-frame & 6.2\% & -2.2\% \\
SDSS 1254 & 2014-03-09 & 74 & 14 & clear\tablenotemark{f} & 4.27/2.90 & multi-frame\tablenotemark{g} & 4.8\% & +1.7\% \\
SDSS 1254 & 2014-03-12 & 63 & 14 & clear & 2.14/1.72 & multi-frame\tablenotemark{g} & 1.8\% & -0.9\% \\
2M 1324 & 2014-03-13 & 104 & 31 & some clouds & 1.83/1.65 & none\tablenotemark{g} & 5.4\% & -0.2\% \\
2M 1324 & 2014-03-14 & 82 & 31 & clear & 2.14/2.09 & multi-frame\tablenotemark{g} & 4.9\% & +1.1\% \\
2M 1503 & 2014-05-17 & 21 & 13 & many clouds & 1.53/1.45 & multi-frame\tablenotemark{h} & 1.7\% & -2.1\% \\
2M 1503 & 2014-05-18 & 54 & 13 & some clouds & 1.83/1.78 & none\tablenotemark{g} & 1.6\% & +0.3\% \\
WISE 1506 & 2014-05-19 & 53 & 22 & some clouds & 2.14/2.18 & multi-frame & 2.0\% & +1.2\% \\
WISE 1506 & 2014-05-20 & 35 & 22 & some clouds & 2.14/1.83 & single-frame & 1.5\% & -1.7\% \\
WISE 1506 & 2014-06-14 & 38 & 15 & clear & 1.83/1.66 & multi-frame & 1.8\% & NA\tablenotemark{i} \\
SDSS 1520 & 2014-05-21 & 70 & 26 & clear & 2.44/1.67 & multi-frame & 2.3\% & +1.5\% \\
SDSS 1520 & 2014-06-14 & 42 & 26 & clear & 1.83/1.50 & single-frame & 2.1\% & -2.2\% \\
SDSS 1624 & 2014-06-15 & 34 & 52 & clear & 1.53/1.63 & none & 4.4\% & +6.8\% \\
SDSS 1624 & 2014-06-16 & 26 & 52 & some clouds & 1.53/1.60 & none & 6.8\% & +1.2\% \\
SDSS 1624 & 2014-06-17 & 43 & 52 & few clouds & 1.22/1.41 & single-frame & 2.5\% & -0.6\% \\
SDSS 1624 & 2014-06-18 & 39 & 52 & clear & 1.53/1.50 & single-frame & 4.2\% & -1.2\% \\
2M 1629 & 2014-06-15 & 38 & 49 & clear & 1.83/1.58 & single-frame & 2.2\% & +8.4\% \\
2M 1629 & 2014-06-16 & 28 & 49 & some clouds & 1.53/1.32 & none & 1.5\% & -0.4\% \\
2M 1629 & 2014-06-17 & 33 & 49 & few clouds & 1.53/1.38 & single-frame & 1.7\% & -1.0\% \\
2M 1629 & 2014-06-18 & 40 & 49 & clear & 2.14/1.66 & multi-frame & 2.3\% & +0.2\% \\

\enddata
\tablenotetext{a}{Images taken with 200 second exposures in the f814w filter and used in the final analysis.}
\tablenotetext{b}{The number of objects for which photometry was acquired, including the brown dwarf target and reference field stars.}
\tablenotetext{c}{Diameter of the optimal photometric aperture, and median seeing, both in arcseconds (see Section \ref{sec:optphot}).}
\tablenotetext{d}{RMS error of normalized photometry, or equivalently the standard deviation of the lightcurve: $\sigma$ as defined in Section \ref{sec:photres} and used in equation \ref{eq:MADmet}.}
\tablenotetext{e}{The offset of the mean for this nights' measurements from the overall mean of all our photometry for this object.}
\tablenotetext{f}{The seeing was very bad on this night.}
\tablenotetext{g}{For this data set, the sky-subtraction method was optimized between
multi-frame sky subtraction and no sky subtraction; single-frame sky subtraction was not tested.}
\tablenotetext{h}{For this data set, multi-frame sky subtraction was adopted
without testing the photometric precision of single-frame or of no sky subtraction}
\tablenotetext{i}{The offset of the nightly mean for this data set cannot be
assigned, because (due to a pointing error) the set of reference stars was different from the set used for the
other nights' observations of WISE 1506.}
\end{deluxetable}

\subsection{Image Processing} \label{sec:improc}
We trim the images to remove an unusable overscan region and edge-regions that
show vignetting and other unusual behavior; zero-subtract them using a median stack
of zero images; and then flatfield them using a normalized median stack of dome
flats\footnote{Established wisdom at Kitt Peak states that sky flats from the 2.1m do
not produce good results.}.  We interpolate
over a few small regions of bad pixels, including two partial bad columns.

Our further processing is more specific to the relatively long optical wavelengths at
which we observe and the specific goals of our analysis.  We perform preliminary
photometry on bright unsaturated stars, and reject any images with more than
50\% extinction from telluric clouds. Most remaining images have much less
than 50\% extinction.  Temporary copies of the remaining images are made
and normalized to a mean sky background value of 1.0.
The normalized frames for each object on each night are median-stacked to produce
a sky frame for that object and night.  Only faint residual halos
of bright stars remain in this sky frame, thanks to the large
number of non-redundant dithers.  As is typical of CCD observations at long
wavelengths, the sky frames show a complex pattern of
interference fringes caused by the monochromatic light from telluric night-sky
emission lines internally reflecting within the thickness of the silicon
detector.  We subtract a scaled version of a master sky frame from each science frame.
We refer to this type of processing as single-frame sky subtraction.

Single-frame sky-subtracted images are cosmetically cleaner than
pre-subtraction, but in many cases, sky-related artifacts can
still be seen, especially when we shift, rotate, and stack the images to create a deep, astrometrically 
registered master star image. In such an image, dark halos exist around bright stars
and galaxies; there are other variations in the background; and the edges of some
of the individual frames that went into the stack can still be seen.  The dark halos are
due to the existence, in the sky frame, of faint residuals from the halos of
bright stars, which survive our median stack despite the standard 5$\sigma$ 
sigma-clipping that we apply.  

The background variations and image-edge effects in the master star image
show that the spatial brightness distribution of the sky background
has changed over the course of the night.
Such change is not surprising: the sky background is composed of several
components, including fringing from night-sky emission lines, moonlight in the sky, 
and scattered moonlight in the dome.  The relative contributions of these
different components change over time, while telluric clouds, when present,
cause further variations. A single sky frame constructed for the whole night is
therefore only a crude approximation of the true sky background on any individual frame.
To do better, we use a technique we initially developed for use on
infrared ($J$-band) brown dwarf monitoring data, where it greatly improved
results.

To track variations in the sky background, we split up the science images into
several sets, produce a different sky frame from each set, and use that frame
for sky subtraction only of images within its set.  By itself, however, this
procedure would make the problem of residual halos from bright stars
in the sky frames much worse, because of the smaller numbers of images being
stacked.  To avoid this, we mask the stars --- not simply their bright cores, which
vanish in the median combine anyway, but their faint diffuse halos, which persist
in dithered stacks almost regardless of what scheme of outlier rejection is employed.

We construct the mask based on the stacked, master star image, which allows
us to mask even diffuse sources too faint to measure on individual frames.
We mask all pixels above a given threshold (e.g. 20 ADU), but we also
mask the haloes of bright stars even at brightnesses below this value by
masking all pixels that lie within given radii of pixels brighter than
certain empirically determined thresholds: for example, points within 25 pixels of any pixel brighter than
50000 ADU; or within 10 pixels of a point brighter than 2000 ADU, etc.
We apply the mask to each original science frame by reversing the shift and rotation that was required to
register that frame with the master star image from which the mask was made. 
We normalized the masked science frames and stack them to produce
typically three to five sky-subtraction frames capturing the evolution in the sky
background distribution over the night.  We subtract scaled versions
of these new sky images from the corresponding original science frames.  We
refer to this type of processing as multi-frame sky subtraction.

Multi-frame sky subtraction always improves the appearance of the final stacked
images.  However (in contrast to our earlier experience with $J$-band photometry)
we find that the improvements it yields in the final photometric scatter are
typically modest, and that our results would suffer only slightly if we had not
employed it.  In some cases, indeed, the photometric scatter turns out to be
lower in the single-frame sky subtracted or even the unsubtracted data.
Possible explanations for this include rapid cloud-induced variations in
the background; a target close to a bright star whose diffraction
rays create a complex illumination pattern that cannot be fully masked;
and increased Poisson noise in the multiple sky frames
produced by stacking smaller numbers of images.  In any case, we proceed with our analysis
using only the lowest-scatter photometry we have, whether this is obtained from
images processed with multi-frame, single-frame, or no sky subtraction.

\section{Photometric Analysis and the Identification of Variables} \label{sec:phot}

The objective of our analysis in this section is to identify any T dwarfs that
exhibit statistically significant variability that is not a product of instrumental systematics,
and to characterize such variability in a phenomenological sense.  
At present we will assume that non-instrumental variability is 
astrophysical and intrinsic to the brown dwarfs.  However,
in Section \ref{sec:vars} we will consider (and rule out) the possibility
that the variations could be caused instead by chromatic extinction in the Earth's atmosphere.  
Our reason for postponing the consideration of telluric effects to Section \ref{sec:vars} is
that our arguments there will depend on the detailed photometric characterization developed
in the present section.

As with our image processing,
our photometric analysis deals first with individual nights. In Section \ref{sec:indiv} 
we analyze the photometry of each T dwarf on each given night independently
of other nights' observations of the same target.
This is appropriate because
the variability amplitude of Luhman 16B is known to change greatly from one
night to the next.  However, since not all variable brown dwarfs share this behavior,
and longer-period variations might manifest themselves partly as changes in the nightly mean flux,
we also perform a self-consistent analysis of all the photometry of each given object
in terms of its nightly means in Section \ref{sec:multi-night}.

\subsection{Analyzing the Photometry from Individual Nights} \label{sec:indiv}

\subsubsection{Optimized Relative Photometry} \label{sec:optphot}
We use aperture photometry to measure typically 15 to 70 stars (see Table \ref{tab:data} for exact numbers)
in addition to the brown dwarf target on each image.  Requirements for these stars
are that they be unsaturated and well within the detector's linear regime; mostly
brighter than the target but ranging down to 50-70\%
of the target's flux; visible on every science frame; distributed approximately symmetrically
around the target; and not close to the edge of the field. 

We construct relative photometry of the brown dwarf by taking the ratio of
its measured flux on each image to the summed fluxes of all the
reference stars measured on the same image.  We also construct relative
photometry of each star by taking the ratio of its flux to the summed
flux of all the other stars:

\begin{equation} \label{eq:relphot}
R_{ij} = \frac{F_{ij}}{\sum_{k \neq j} F_{ik}}
\end{equation}

\noindent where $F_{ij}$ is the measured raw flux of Star $j$ on image $i$.
The resulting photometric time series
obtained for each object is normalized to a mean of 1.0. We screen the stars
for variability by plotting the standard deviations of their 
normalized photometry as a function of their mean fluxes on
a log-log scale.  Significant positive outliers from the resulting linear trend
are rejected from the reference set. Low-level variables could pass
this screening, but this is acceptable because our use of many
reference stars, including some much brighter than our
targets and thus measurable with higher precision, ensures that
such undetected variables can have only a negligible effect on the
final relative photometry of our targets.

The optimal photometric aperture is determined by minimizing
the average standard deviation in normalized photometry of stars with
brightnesses similar to that of our brown dwarf target.  This same
criterion is also used in the optimization of the sky subtraction
method that was discussed above.  The aperture optimization is carried
out over aperture diameters ranging from 4 to 14 pixels in 1 pixel increments.
The same optimal aperture is used for all stars in the 
field of a given science target on a given
night, but optimal apertures are re-calculated from night to night and
from one science target to another.  The optimal photometric aperture
is effectively set by a tradeoff between systematic and random errors:
increasing the aperture reduces systematic effects due to aperture losses,
which can in principle affect relative photometry if the point spread function
varies across the field.  Decreasing the photometric aperture
reduces the effect of background Poisson noise. Optimal apertures tend
to be smaller on nights of consistent good seeing, and for very faint objects
where background noise dominates the photometric error budget. For
our three faintest targets, the diameter of the aperture is, on average,
1.01 times the median full width at half maximum (FWHM) of the seeing disk.
For the rest of our sample, by contrast, the optimal aperture averages 1.16 times
larger than the FWHM.

Our photometric precision, as determined by the RMS error of normalized
relative photometry (i.e. the standard deviations of our lightcurves), varies with target flux and seeing.  Magnitudes
for all our targets are given in Table \ref{tab:targs}, and seeing
values and per-image RMS errors are given in Table \ref{tab:data}. Our best
precision (RMS error 1.2\%) was obtained for 2M 1106 on March 7.  At magnitude
18.8 in the f814w filter, this is our third brightest target.  Our worst
precision (RMS error 6.8\%) was for SDSS 1624 on June 16: at f814w magnitude
20.3 this is our faintest target. The effect of seeing on precision is most clearly seen
for SDSS 1254 (f814w magnitude 19.1), for which we obtained RMS error
4.8\% on March 9 in three-arcsecond seeing and 1.8\%
on March 12, when the seeing was 1.7 arcsec.

To obtain the absolute photometry given in Table \ref{tab:targs} from 
our measurements using apertures optimized for relative photometry, we apply an aperture correction derived
on a per-image basis from measurements of bright stars, correcting the T dwarf fluxes to an
effective aperture of diameter 40 pixels (12.2 arcsec).  The photometry is then 
calibrated using measurements of \citet{Landolt1992} standard stars with the same
40 pixel aperture.

\begin{figure}
\plottwo{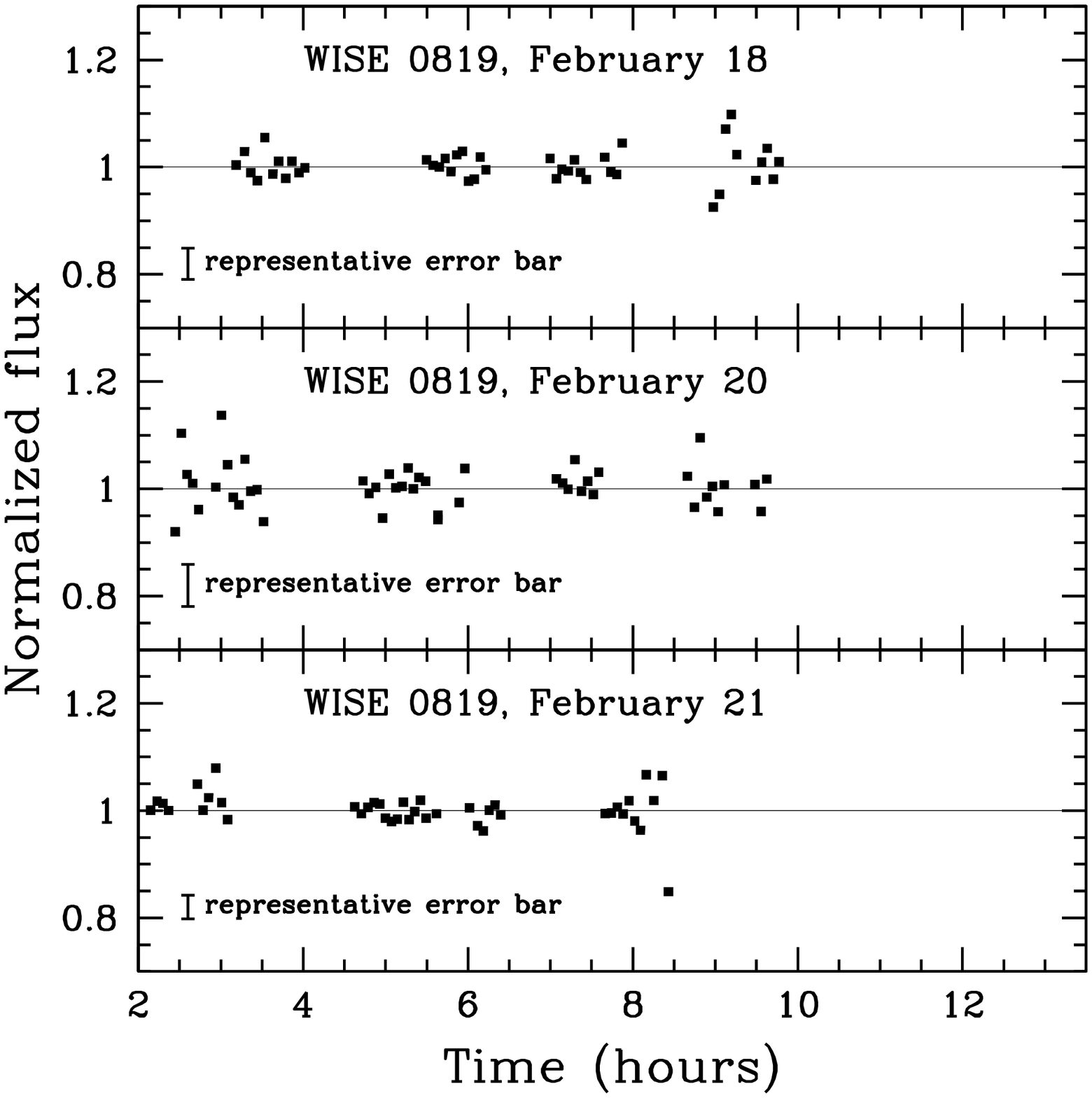}{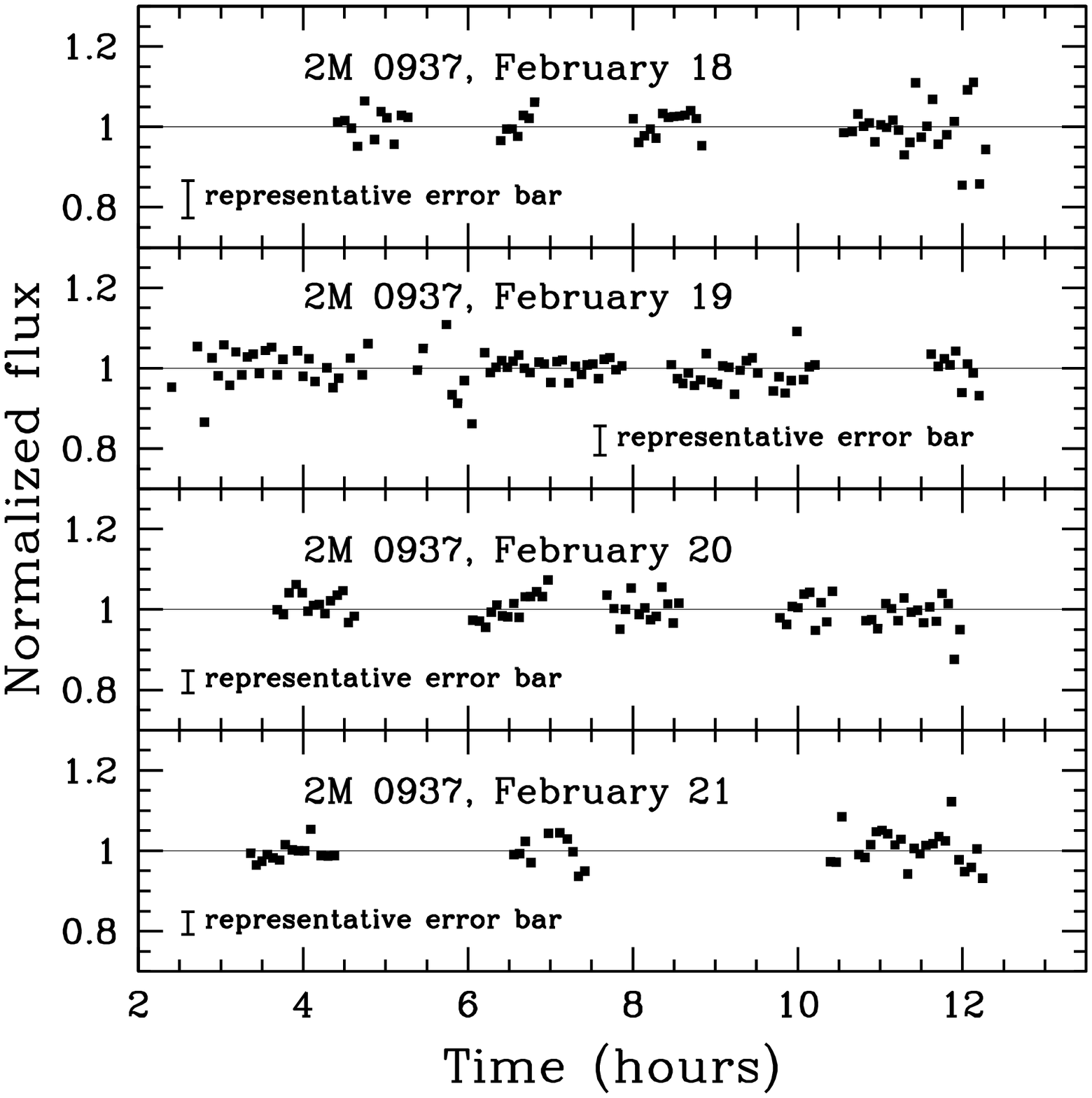}
\caption{Our photometric monitoring data for T dwarfs WISE 0891 
(left) and 2M 0937 (right), processed and presented one night at a time.  
The extent of random scatter in the photometry varies depending on the brightness of the
object and on the quality of the night. A representative error bar
is shown at the lower left or bottom-center in each plot.
\label{fig:nightplot01}}
\end{figure}

\begin{figure}
\plottwo{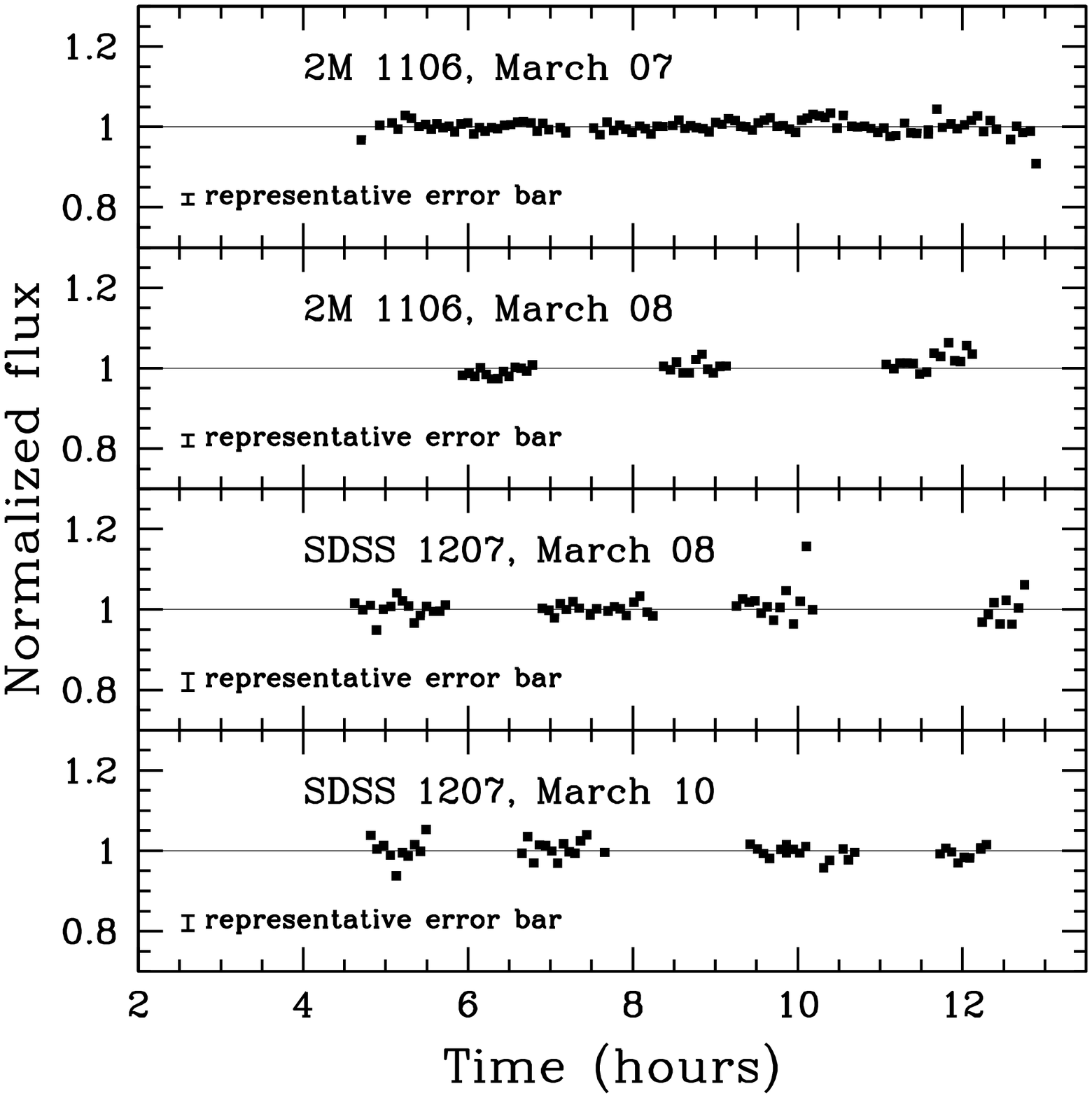}{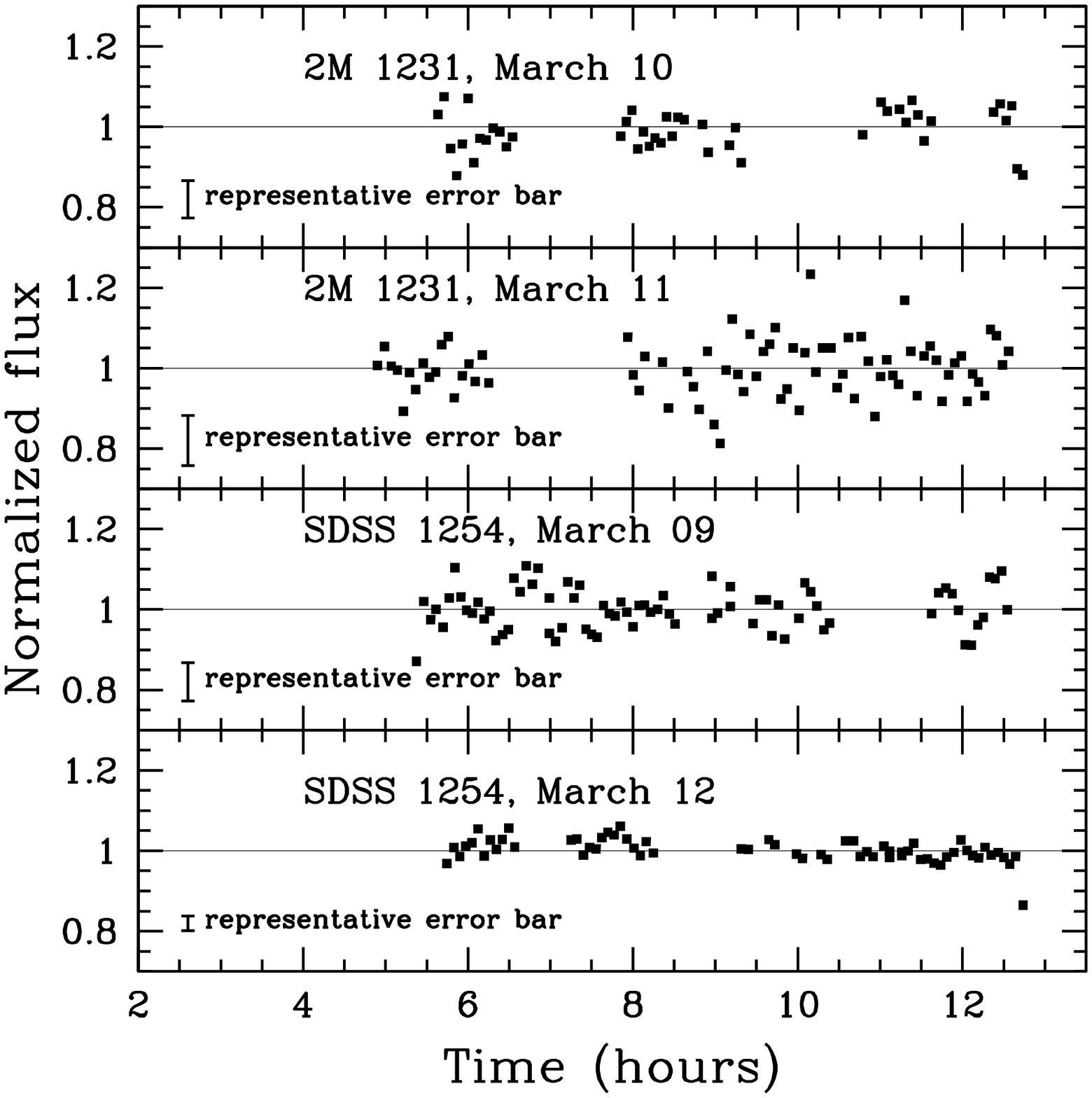}
\caption{Our photometric monitoring data for 2M 1106 and SDSS 1207 (left)
and 2M 1231 and SDSS 1254 (right).    A representative error bar
is shown at the lower left in each plot.
The scales on both axes and the type of
data presented are the same as in Figure \ref{fig:nightplot01}.
\label{fig:nightplot02}}
\end{figure}

\begin{figure}
\plottwo{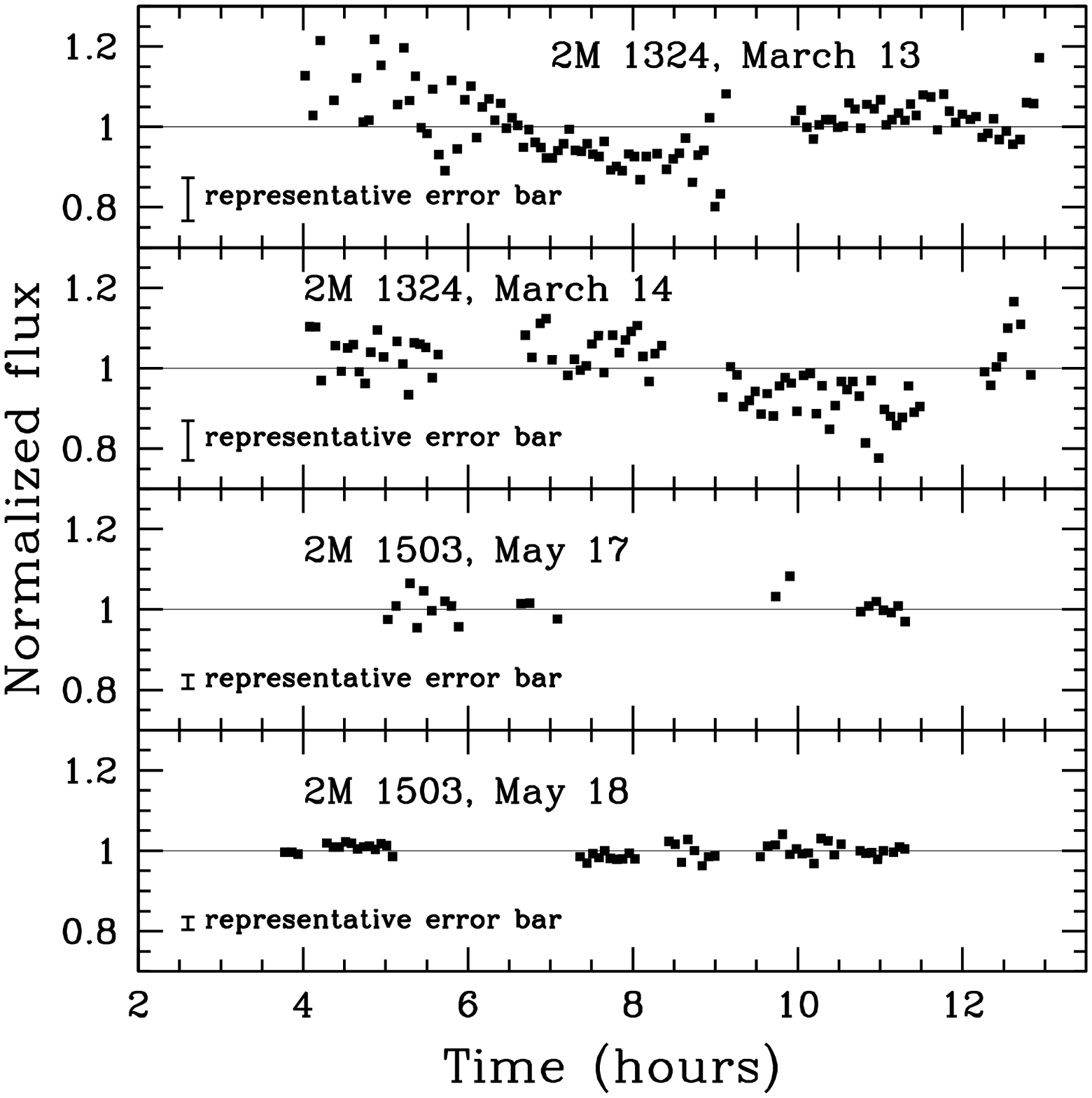}{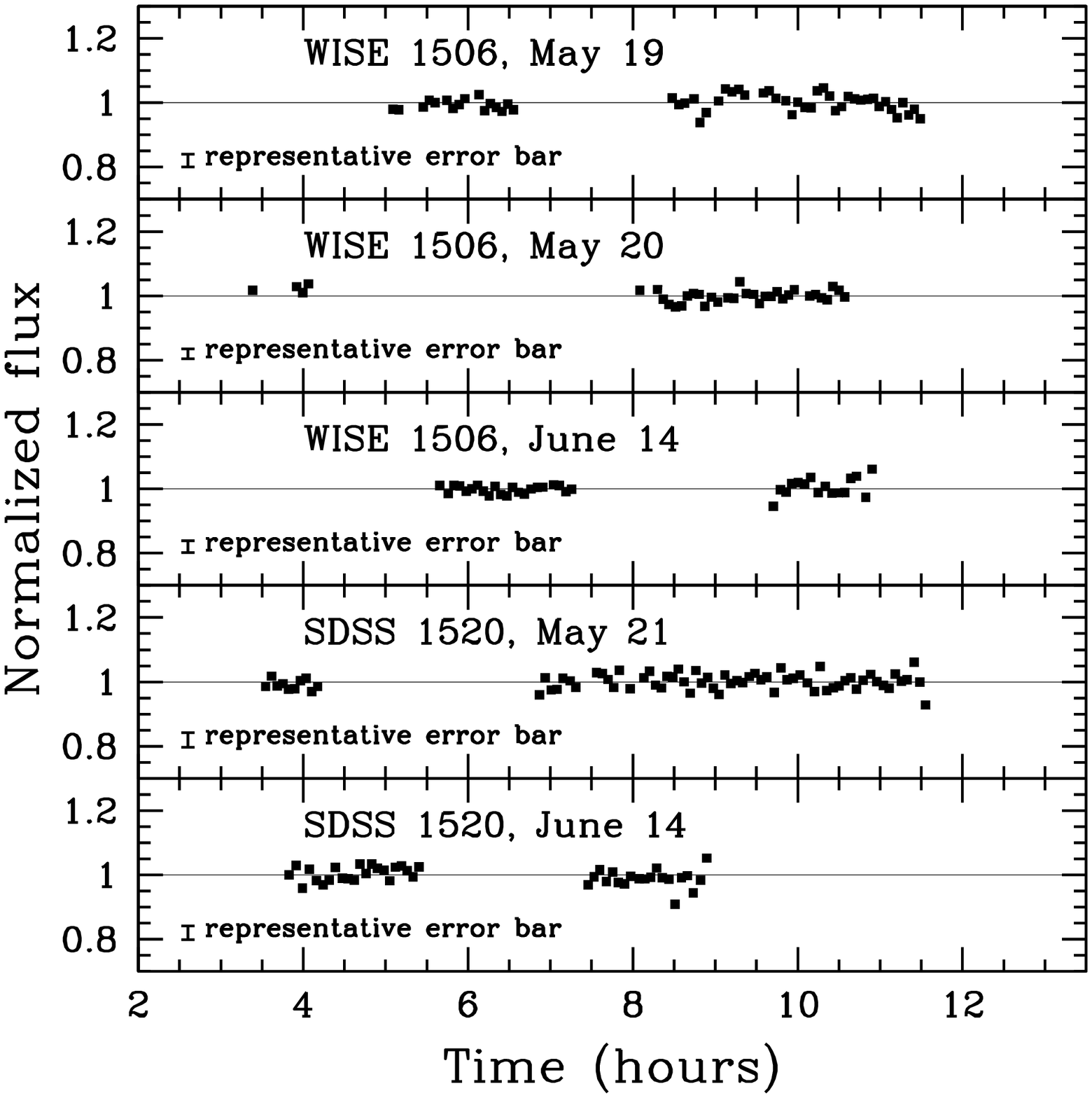}
\caption{Our photometric monitoring data for 2M 1324 and 2M 1503 (left)
and WISE 1506 and SDSS 1520 (right).   A representative error bar
is shown at the lower left in each plot.
We find 2M 1324 to be a high-amplitude 
variable. The scales on both axes and the type of
data presented are the same as in Figures \ref{fig:nightplot01}--\ref{fig:nightplot02}.
\label{fig:nightplot03}}
\end{figure}

\begin{figure}
\plottwo{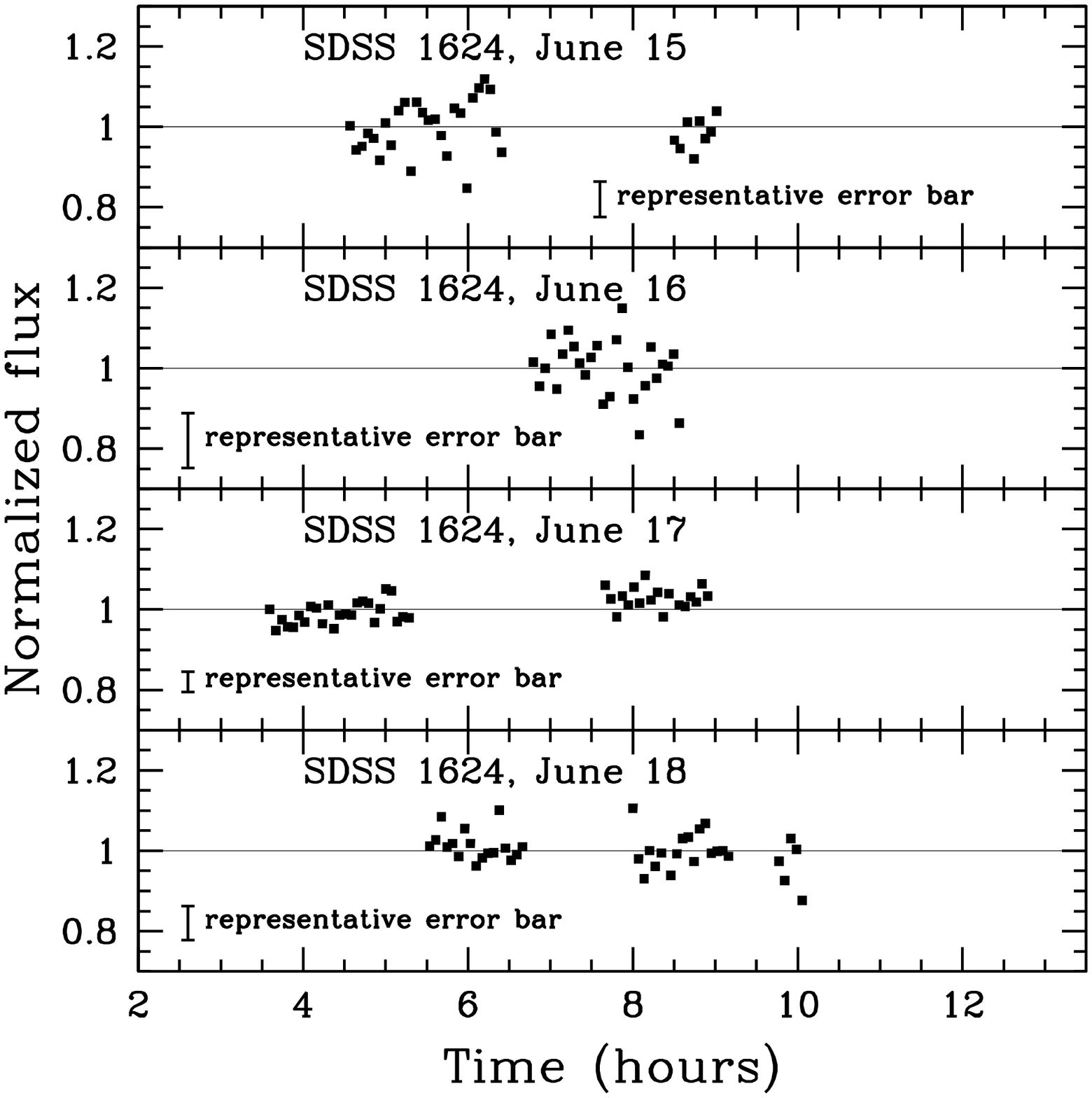}{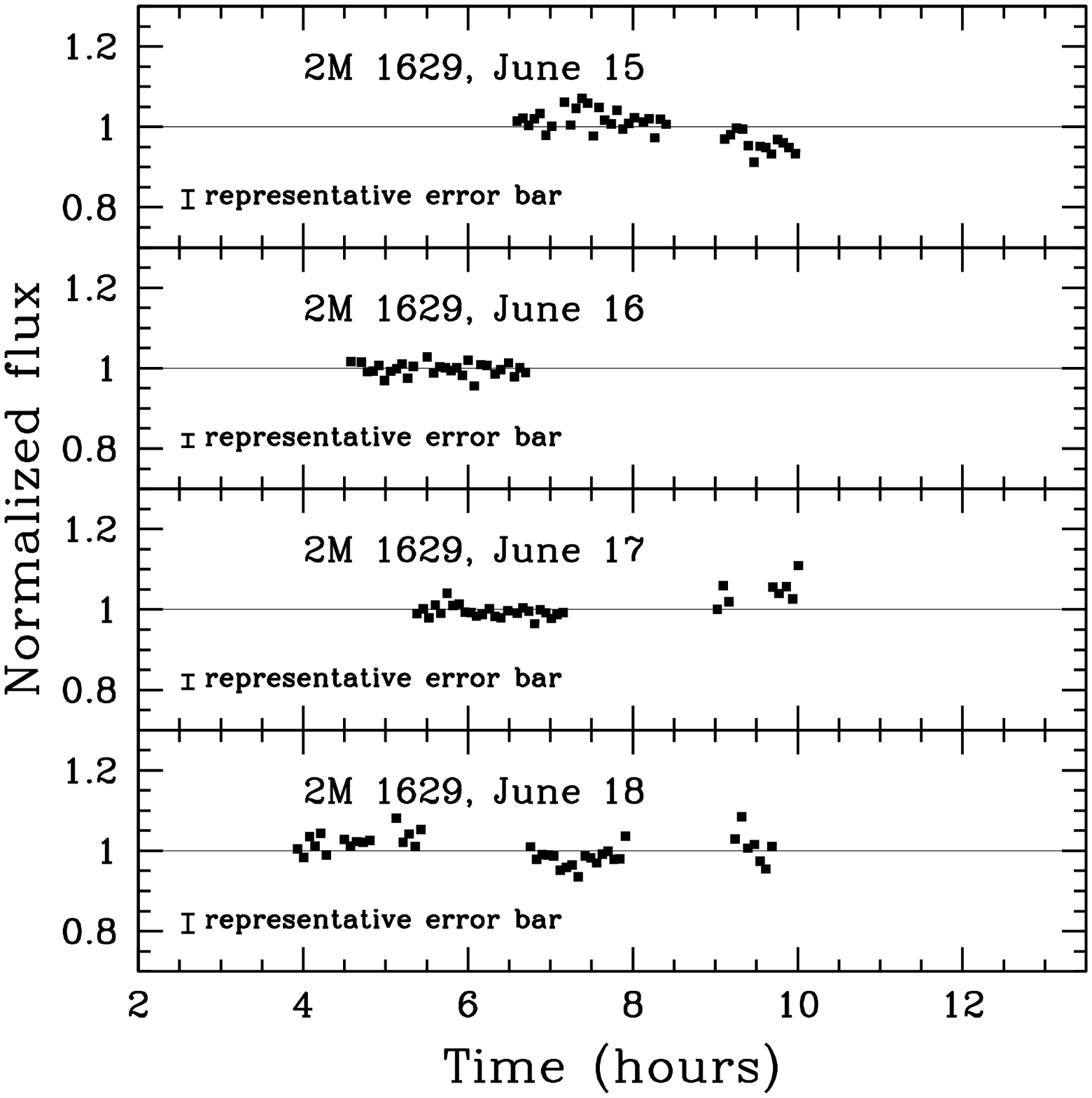}
\caption{Our photometric monitoring data for SDSS 1624 (left) and
2M 1629 (right).  A representative error bar
is shown at the lower left or bottom-center in each plot.
The scales on both axes and the type of
data presented are the same as in Figures \ref{fig:nightplot01}--\ref{fig:nightplot03}.
\label{fig:nightplot04}}
\end{figure}

Figures \ref{fig:nightplot01} through \ref{fig:nightplot04} present the optimized
relative photometry of each of our brown dwarf targets on each night when
it was observed.  These figures strongly suggest that
the T2.5 dwarf 2M 1324 is a high-amplitude variable.  No other target shows
obvious variability at this stage of the analysis.

\subsubsection{Single-Night Variability Metric} \label{sec:photres}

Visual examination of the lightcurves of both stars and brown dwarfs shows
that systematic variations are common in this data set: that is, coherent variations
that are highly significant in the context of the random error are present
in far too many objects for all of them to be reasonably attributed to true
astrophysical variability.  It follows that identifying astrophysical
variables does not reduce to the simple task of determining which objects
have variations that are significant relative to their random error.  This is
a familiar situation when working with photometric monitoring data 
\citep{Koen2013a,Heinze2013,Radigan2014,Metchev2014}. The
solution is to apply a metric for coherent variability, and thereby quantify
the typical systematic effects present in the data using reference stars
that do not appear to be astrophysical variables.  A threshold variability level may
then be defined, above which objects are classified as true astrophysical
variables with, e.g., 95\% confidence \citep{Heinze2013,Radigan2014,Metchev2014}.  

This scheme tends to be inherently
conservative, since it assumes that any variations observed
in field stars that are not obvious astrophysical variables are purely
due to measurement systematics. In reality, low-level astrophysical
variability will certainly also be present in some of the stars, and this
will elevate the variability detection threshold above where it would be
if measurement systematics were the only cause of coherent variations in
the field stars.  This conservatism is a desirable attribute, since
it reduces the chance of false positives in variable identification, and
the attendant scientific misinterpretation.

The optimal variability metric depends on the characteristics
of the available data.  For Spitzer data we have previously used a metric
related to the Lomb-Scargle periodogram power \citep{Heinze2013,Metchev2014},
as did \citet{Radigan2014} for contiguous $J$-band data. The photometric time-series we consider herein
have gaps, uneven sampling, occasional extreme outliers, and widely differing
numbers of valid data points.  While a periodogram-based metric could still
work in principle, it would be difficult to make it acceptably uniform
and reliable.  We choose instead a variability metric based on a smooth functional fit to the
photometry, combined with iterative outlier rejection. A truncated
Fourier series is a natural choice for a functional form to fit rotationally
periodic brown dwarfs, and it will also fit arbitrary non-periodic variations
(whether systematic or astrophysical) provided the period is allowed to range
up to large enough values.  Based on examination
of the single-night relative photometry, we identify a two-term Fourier
series as the most appropriate function:

\begin{equation} \label{eq:fourmod1}
F(t) = a_0 + a_1 \sin \left ( \frac{2 \pi t}{P}  + \phi_1 \right) + a_2 \sin \left ( \frac{4 \pi t}{P} + \phi_2 \right)
\end{equation}

\centerline{or equivalently, for purposes of linear least-squares fitting:}

\begin{equation} \label{eq:fourmod2}
F(t) = a_0 + b_1 \sin \left( \frac{2 \pi t}{P} \right)+ b_2 \cos \left( \frac{2 \pi t}{P} \right)+ b_3 \sin \left( \frac{4 \pi t}{P} \right)+ b_4 \cos \left( \frac{4 \pi t}{P} \right).
\end{equation}

The six parameters to be fit are $a_0$, the $b_i$, and the period $P$.
If the period is fixed, the other parameters can be found using a linear
least-squares fit.  Two iterations of sigma-clipping
with a 3$\sigma$ threshold are employed to reject outliers.  We probe trial periods
ranging from 2.0 to 50.0 hr in 0.01 hr increments\footnote{A few brown dwarfs are known to have
rotation periods shorter than our 2 hr minimum trial period, but since our
Fourier model has two terms, it can fit sinusoidal variability down to half this period.  There is no evidence
for such short-period variability in any of our targets.}, 
and we adopt the period producing the
smallest RMS residual from the least-squares fit. We use this method to fit the
single-night, normalized relative photometry of each brown dwarf and each reference star.

\begin{figure}
\includegraphics[scale=0.8]{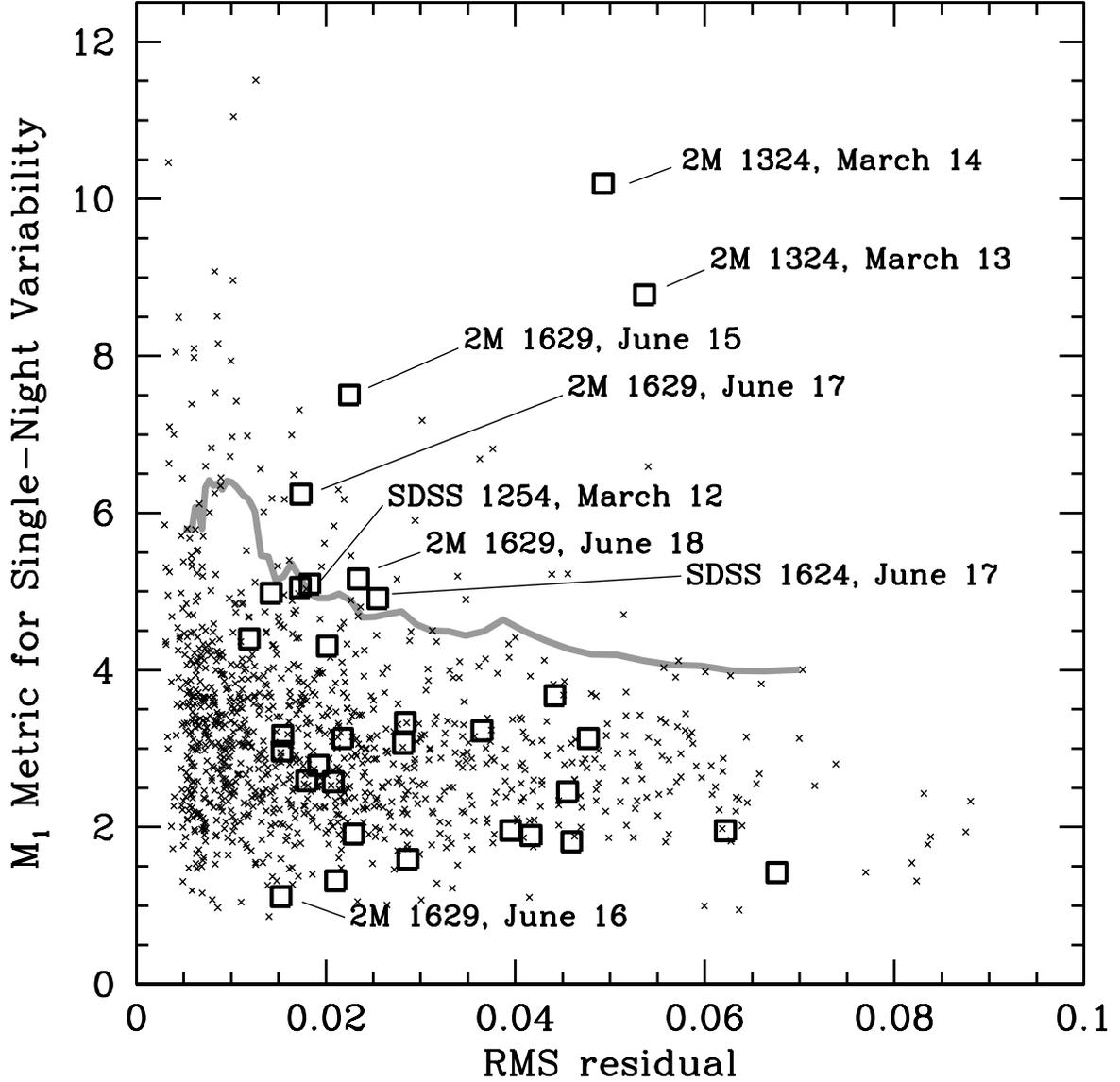}
\caption{The extent of non-random variability for both brown dwarfs
(large squares) and field stars (small x's) in our data within individual
nights, measured using
the metric $M_1$ discussed in Section \ref{sec:photres}.  Each point represents a single object
on a single night: thus there is more than one point corresponding to each
brown dwarf.  The horizontal axis is the RMS residual between the data and the
best-fit model described by Equation \ref{eq:fourmod2}; this is the same value
quoted in the RMS column of Table \ref{tab:data} and given as $\sigma$ in Equation \ref{eq:MADmet}.
Stars with small RMS residuals are
bright and can have systematic variations due to aperture losses that
are too small in fractional terms to affect fainter objects such as
the brown dwarfs.  The gray line is the 95th percentile upper envelope
for the stars.
\label{fig:metric}}
\end{figure}

To calculate our variability metric, we evaluate the resulting model only where it is constrained by
the data: that is, at the times $t_i$ when the actual images were acquired.  We
find that the mean absolute deviation (MAD) of the model points $F(t_i)$ from their own mean is
a good measure of the strength of variations.  To account for differences in
the level of photometric precision and the number $n$ of photometric measurements available
in our different data sets for individual nights, we multiply the MAD by  $\sqrt n/\sigma$, where $\sigma$ is the RMS residual between the
data and the model fit for the object being considered.  Thus our
metric becomes:

\begin{equation} \label{eq:MADmet}
M_1 = \frac{\mathrm{MAD}(F(t_i))}{\sigma} \sqrt n
\end{equation}

\noindent where the MAD is calculated over all the model points for the object in
question on a given night. We apply this metric
to all of our normalized photometric time series of brown dwarfs and of field
stars, and plot the results as a function of $\sigma$ in Figure \ref{fig:metric}.
There is a weak inverse dependence of $M_1$ on $\sigma$.  This is because the photometric apertures we have
used are optimized for objects in the rather faint magnitude range of our photometric
targets: thus bright stars, which have small values of $\sigma$, are measured with
sub-optimal apertures and are more likely to exhibit significant systematic effects.
This does not affect our ability to identify variable brown dwarfs.  As our
threshold for classifying a target as a candidate variable, we choose
the 95th percentile of $M_1$ for the field stars, calculated in a sliding box that
is defined logarithmically such that it always spans a factor of 2.0 in $\sigma$.
This threshold is indicated by a heavy gray line in Figure \ref{fig:metric}.

\subsubsection{Identification of Variables in Individual Nights} \label{sec:indvar}

The analysis described in Section \ref{sec:photres} and illustrated by Figure \ref{fig:metric}
confirms the visual impression from Figure \ref{fig:nightplot03} that the T2.5 dwarf
2M 1324 is a high-amplitude variable.  Its variability is highly significant and consistent
on both March 13 and March 14. No field star in its brightness range exhibits
variations of comparable significance in any of our data sets.  The confidence level of this
detection is thus far greater than the nominal 95\% threshold --- indeed, the variability
is beyond reasonable doubt.

2M 1324 is not the only strong variable detection, however.  The T2 dwarf 2M 1629 
also shows variability above our threshold on three out of the four nights it was
observed.  Although one or two false positives at the 95\% confidence level would be
expected among our 32 object-nights of photometry, the probability that a given 
non-variable object would randomly show variations above the 95\% threshold in three out of four
nights is only $4.8 \times 10^{-4}$.  It is highly unlikely that we would see
a false positive of this type in a survey of 12 objects, only three of which were observed
on four separate nights.  Thus, although its amplitude appears to
be less than that of 2M 1324, we also identify 2M 1629 as a variable.

Two other objects are measured at or above the 95\% threshold: SDSS 1624 on June 17,
and SDSS 1254 on March 12.  These could plausibly be either genuine variables or
false positives, so we retain them as unconfirmed candidate variables as we proceed
with our analysis.

\subsection{Analyzing Photometry Across Multiple Nights} \label{sec:multi-night}

The set of reference stars we use for each T dwarf is kept constant from
night to night (except for the June 14 observations of WISE 1506, when a
telescope pointing error prevented this).  This consistency of 
reference star sets allows us to
extend the construction of relative photometry, described by Equation \ref{eq:relphot}, 
to span multiple nights.  The resulting photometric time
series contains all of our photometric data for a given object, and can be
normalized as a unit so that information about possible changes in the nightly
mean flux is not lost.  

\begin{figure}
\plottwo{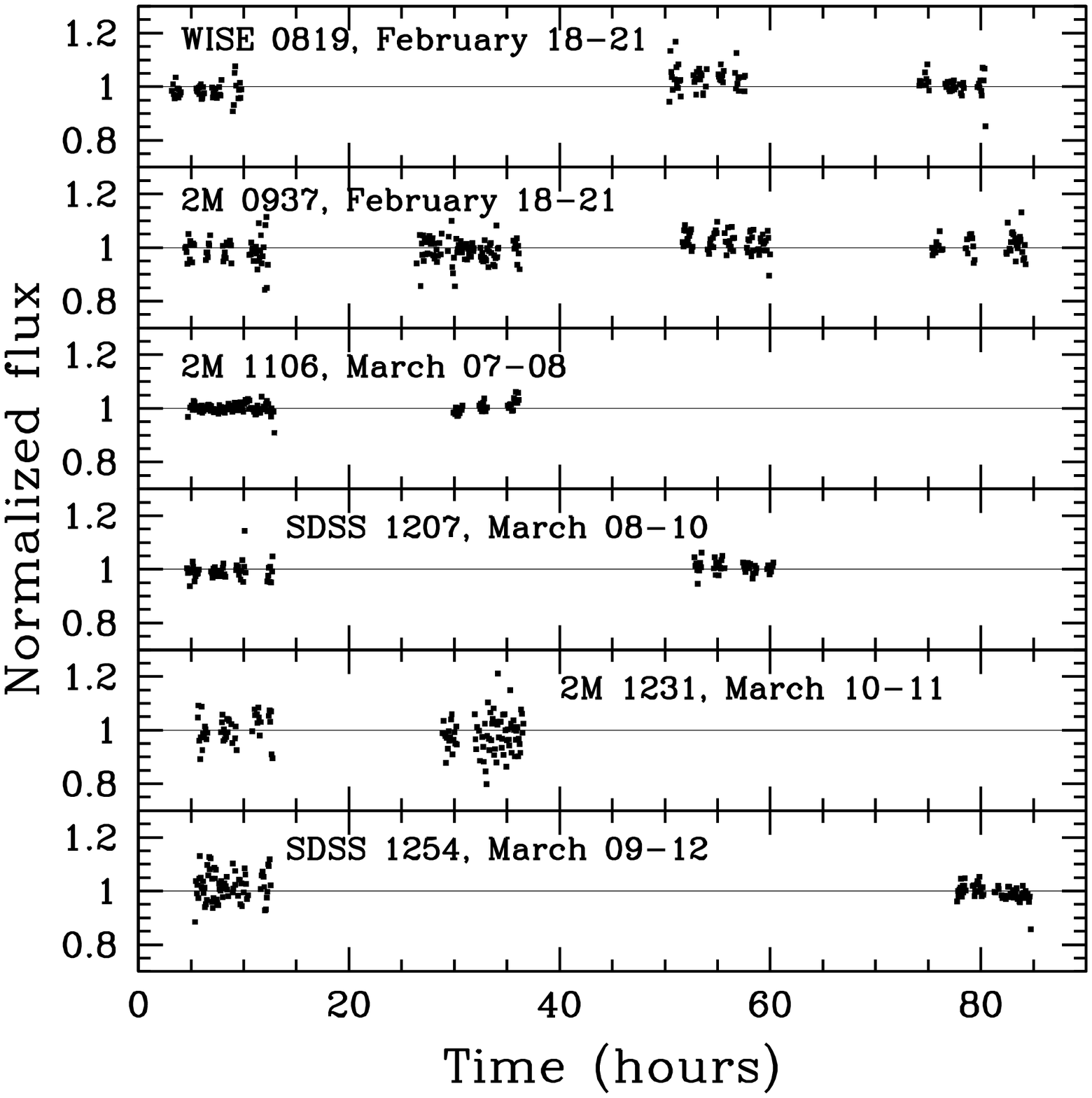}{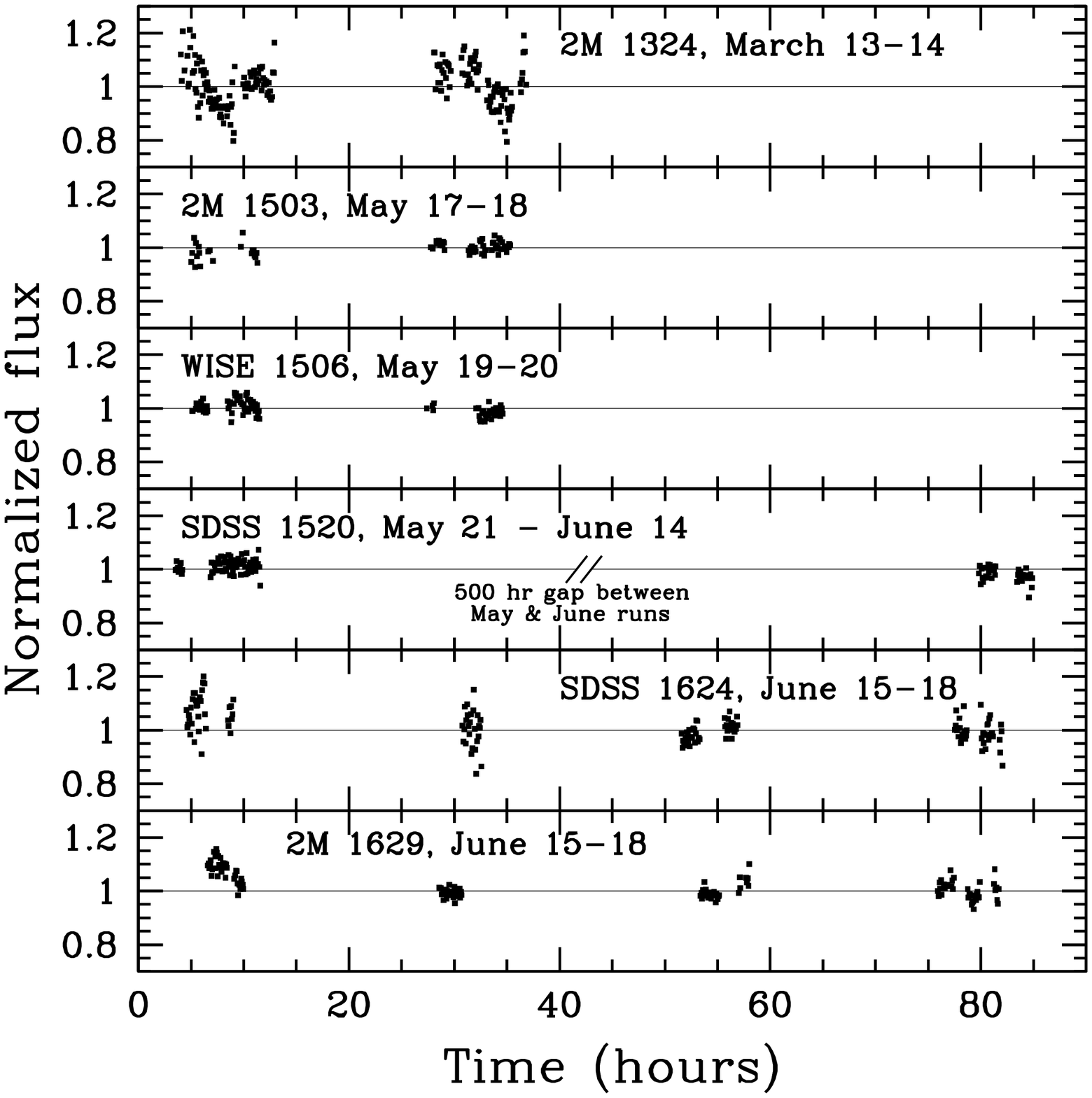}
\caption{Our photometric monitoring data for all targets, 
presented as time series spanning multiple nights.  
A consistent set of reference stars is used for all nights for
each given object, and a single normalization is carried out for the entire time series:
the relative photometry plotted here therefore allows us to probe for astrophysically
real variations in brightness from night to night. Note that because time is
measured from 00:00 UT on the first night that each object was observed,
the different plotted data sets do not all have a common
origin in time.
\label{fig:objects01}}
\end{figure}

We plot this multi-night normalized photometry for each
of our T dwarf targets in Figure \ref{fig:objects01}.  The June 14 photometry
of WISE 1506 is omitted for the reason already given.  Figure
\ref{fig:objects01} shows that 2M 1324's strong
variations center around a mean that changes little from March 13 to March 14.
2M 1629, on the other hand, shows a higher mean brightness on June 15
than on any of the other nights.

\subsubsection{Night-to-Night Variability Metric} \label{sec:nnmet}

We quantify the significance of night-to-night variability using a similar
metric to the one developed in Section \ref{sec:photres}, except that instead
of the MAD of a functional fit, we use the offset of the median value in each
given night from the trimmed mean taken over the entire data set with 10\% rejection\footnote{In our implementation the \textit{trimmed mean} is calculated iteratively, with the single most deviant point rejected after each iteration, until a pre-set fraction (10\% here) of the total points has been rejected.}.  These offsets are given in the last
column of Table \ref{tab:data}.  They (like the offsets of the individual
data points plotted in Figure \ref{fig:objects01}) do not necessarily
average to zero over all nights because of the use of the trimmed mean and because all nights do not have the same number of
data points.  We obtain our metric for variations in the nightly means, which we
refer to as $M_N$, by taking the absolute value of the
offset from Table \ref{tab:data} and scaling by $\sqrt n / \sigma$, where just as
in Section \ref{sec:photres}, $n$ is the number of photometric measurements of a given target
on a given individual night.  
Three relatively faint field `stars' that showed extremely high values of this metric were found
upon close examination to be compact galaxies. Their variations were therefore spurious,
and resulted from their measured fluxes' having a different dependence on seeing and on photometric aperture
than those of true stars. They were removed both from the current analysis and the one
described in Section \ref{sec:indvar}. 

\begin{figure}
\includegraphics[scale=0.8]{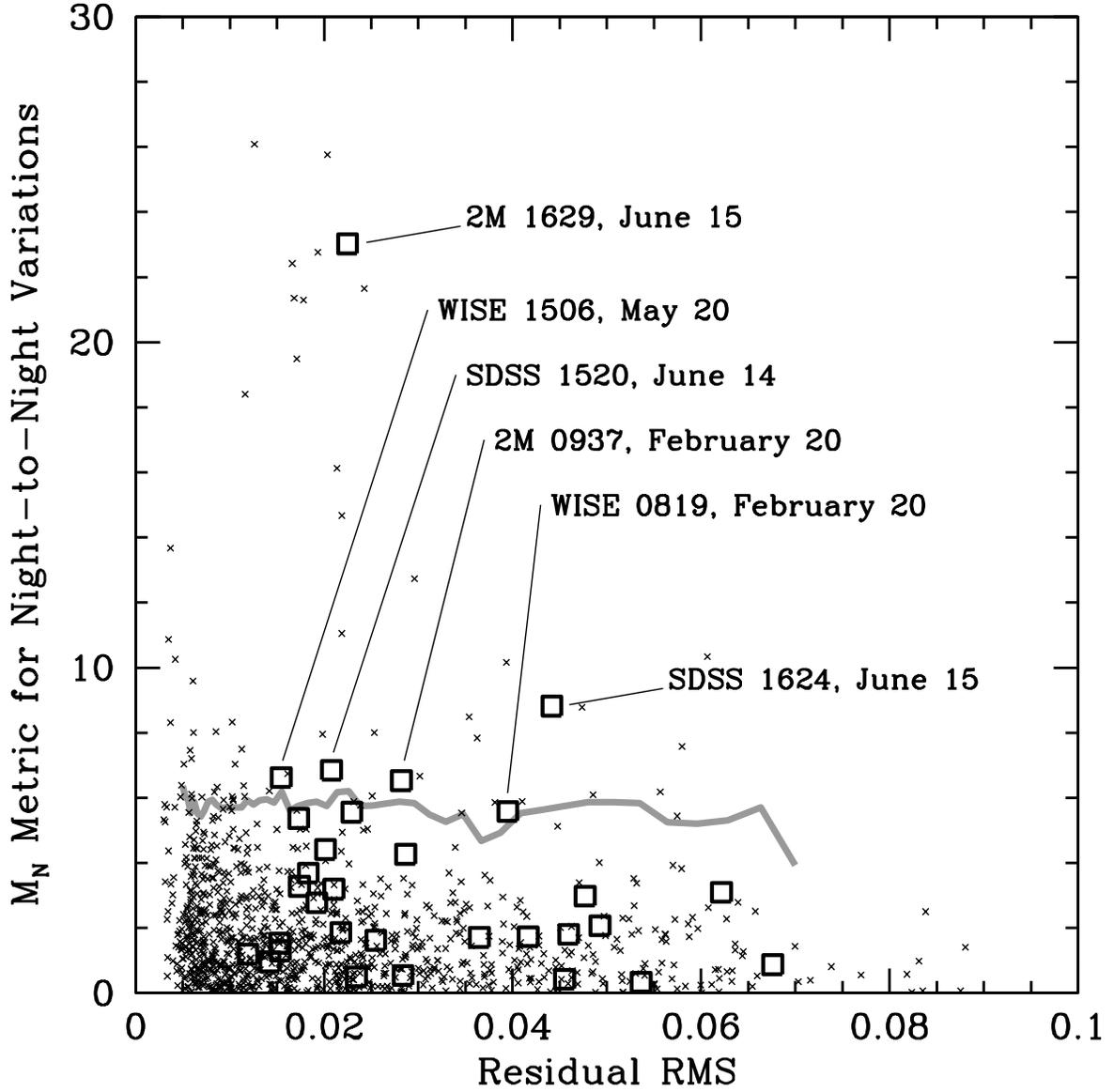}
\caption{The extent of longer-term variability for both brown dwarfs
(large squares) and field stars (small x's), measured based on the extent
to which the mean flux measured on an individual night differs from the
mean over all nights.  The gray line is the 95th percentile upper envelope
for the stars.  2M 1629 appears variable not only in individual nights
but also from one night to the next.  Tentative indications of variability
emerge in other objects, but we cannot confidently identify any of them
as true variables.
\label{fig:nightmean}}
\end{figure}

The results for brown dwarf targets and field stars, after the removal of the spurious variables,
are plotted in Figure \ref{fig:nightmean}. We defer the consideration of whether weaker
seeing correlations could bias the photometry of our T dwarf targets until
Section \ref{sec:seeing}, where we will ultimately conclude that any effects are too small to be relevant
for our analysis.

\subsubsection{Identification of Longer-Term (Night-to-Night) Variables} \label{sec:nnvar}

Figure \ref{fig:nightmean} confirms what was already suggested by Figure \ref{fig:objects01}:
the mean flux of 2M 1629  over the night of June 15 is offset from the overall
mean for this object to a highly significant extent. Since night-to-night
variability is substantially independent of variability within a night (as evidenced
by 2M 1324), this constitutes another piece of evidence confirming 2M 1629 as a genuine
astrophysical variable.  The probability of 4 out of 8 measurements of a given object
spuriously yielding a 95\% confidence detection is even lower ($3.7 \times 10^{-4}$)
than the probability for 3 out of 4 such measurements given already in Section \ref{sec:indvar}.
Thus, even without taking into account the fact that some of the measurements for
2M 1629 are far above the 95\% threshold, it may be identified as a variable with
very high confidence.

The T6 dwarf SDSS 1624 also appears to show significant night-to-night variability.
This object exhibited single-night variability slightly above the 95\% threshold on
June 17 (see Figure \ref{fig:metric}).  Like 2M 1629, SDSS 1624 was also measured on
four nights.  Thus, the two detections of variability above the 95\% threshold, one for single-night
variability and one for variation in the nightly mean, were obtained over eight total measurements.  The
probability of this happening at random for a given object is 5.7\%, and thus one such
event in our survey is not particularly unlikely. Therefore, although the results for SDSS 1624
suggest that it is variable, in our conservative evaluation they not sufficient to confirm it.  The
elevated nightly mean offsets on one night each for WISE 0819, 2M 0937, WISE 1506, and SDSS 1520
have less significance and  do not constitute strong evidence of variability for these objects;
and the case is the same for the suspected single-night variability identified for SDSS 1254 in
Section \ref{sec:indvar}. However, the aggregate presence of
so many detections above the 95\% threshold in both single-night and night-to-night analyses
suggests that variability is more common in T dwarfs than in field stars.

\section{Analyzing the Variability in our T dwarfs} \label{sec:vars}

We have identified 2M 1324 and 2M 1629 as variable brown dwarfs.  
We now characterize the lightcurves of these two objects in more detail,
and set limits on the behavior of the remaining ten objects for which we did
not detect convincing variability.  First, however, we address the issue of
telluric atmospheric effects, and analyze the extent to which these may
have affected our survey.  

\subsection{Telluric Effects in Photometry} \label{sec:telluric}

The two objects we find to be variable are the only
objects in our sample that have been identified as variables in previous
surveys --- a fact which strongly suggests that the variability we detect is intrinsic
to these objects and is not due to telluric effects.  However, our data contain an
apparent coincidence that has led us to analyze the possible influence of wavelength-dependent
telluric extinction. 
This coincidence is seen in the offsets of our targets' nightly mean fluxes, which
are tabulated in Table \ref{tab:data} and plotted in Figure \ref{fig:nightmean}. On two 
of the nights when
we alternated observations between two different brown dwarfs, both objects had nightly
means significantly brighter than the average for either object.  On February 20, both WISE 0819 and
2M 0937 were brighter than their overall means by 3.2\% and 2.3\%, respectively; while on June 15,
both SDSS 1624 and 2M 1629 had respective positive offsets of 6.8\% and 8.4\%. These observations
could be due to random effects and/or astrophysical variability, but only if we accept
a rather improbable coincidence.  Alternatively, the observations could be explained
by a telluric effect specific to those two nights, causing a spurious increase in the measured
fluxes of brown dwarfs relative to reference stars.  The June 15 observation is of particular importance since we
have identified 2M 1629 as a variable, and our final interpretation of its behavior will depend 
on whether its elevated nightly mean on that date is telluric or astrophysical.

\subsubsection{Airmass Effects}

Since we have monitored our objects using relative, not absolute, photometry, achromatic
telluric effects cancel out.  In particular, we can confirm that clouds
causing up to 50\% total extinction do not have a detectable systematic effect on relative
photometry, since we find that acceptable photometry can be obtained through thin clouds
(compare Table \ref{tab:data} and Figures \ref{fig:nightplot01}--\ref{fig:nightplot04}),
and that the offsets of our nightly means are not correlated with the presence or absence
of clouds (see Table \ref{tab:data}). The only telluric effects that would not
cancel would be any due to wavelength-dependence of the telluric extinction across 
the passband of the f814w filter. Evidence for chromatic effects in
$I$ band photometry, with amplitudes of up to 3\% on some nights, has been reported by 
\citet{Bailer01} and \citet{Koen03}.  Such extinction will affect our photometry because the flux of our T dwarf targets
is mostly longward of the filter's central wavelength, while the opposite is true
of typical field stars, as illustrated in the bottom left panel in Figure \ref{fig:airmass}.

We can constrain the likely importance of such effects in our data by probing for
any dependence of our relative photometry on airmass.  If there is significantly
more extinction toward the long-wavelength edge of the filter, the brown dwarfs
should appear fainter at high airmass, while the reverse should be true if
there is more extinction at shorter wavelengths. To probe for such an effect
with high sensitivity, we combine all of our photometric time series (after
normalization to unit median flux for measurements in the airmass range 1.3--1.5),
and plot the data as a function of airmass in the right-hand panel of Figure \ref{fig:airmass}.
No significant dependence is seen.  The exact slope of the linear fit is sensitively
dependent on the scheme of outlier rejection, but we adopt a value of -0.006$\pm$0.005:
that is, the brown dwarfs appeared an average of 0.6\% fainter when observed at two
airmasses vs. one airmass.  The effect is therefore not significant in the context
of our survey --- unless there are isolated nights on which it is much stronger
than average.  We will now investigate this possibility.

\subsubsection{Water Vapor Effects}

The red side of the f814w passband contains strong water absorption lines from 900-960 nm.  
The strength of these lines depends on the amount of precipitable water vapor (PWV)
in the air column through which the telescope is looking.   PWV at Kitt Peak typically
ranges from 2--5mm in February and March, 3--9mm in May, and 2--15mm in June \citep{Wallace1984}.

\begin{figure}
\plottwo{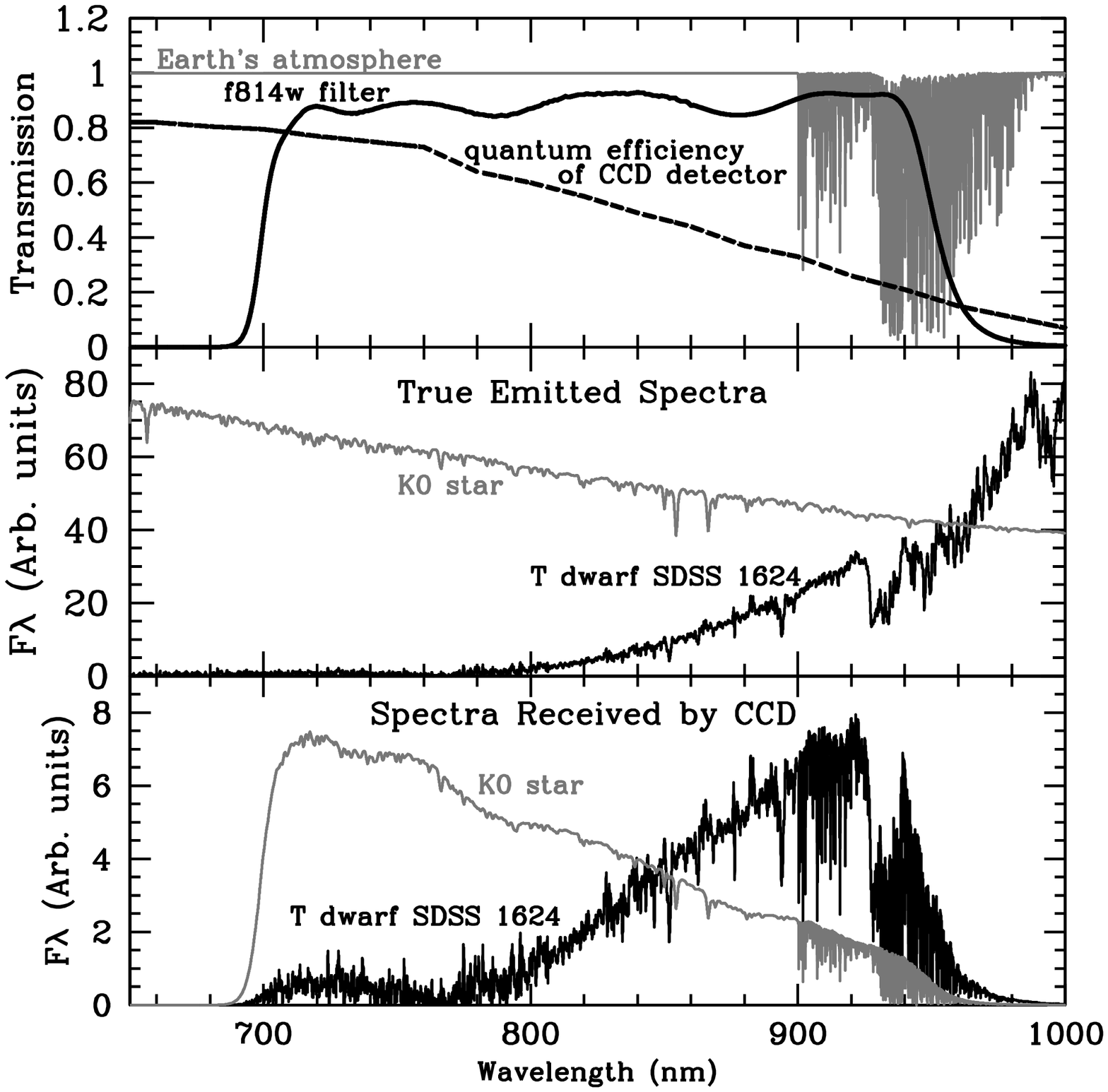}{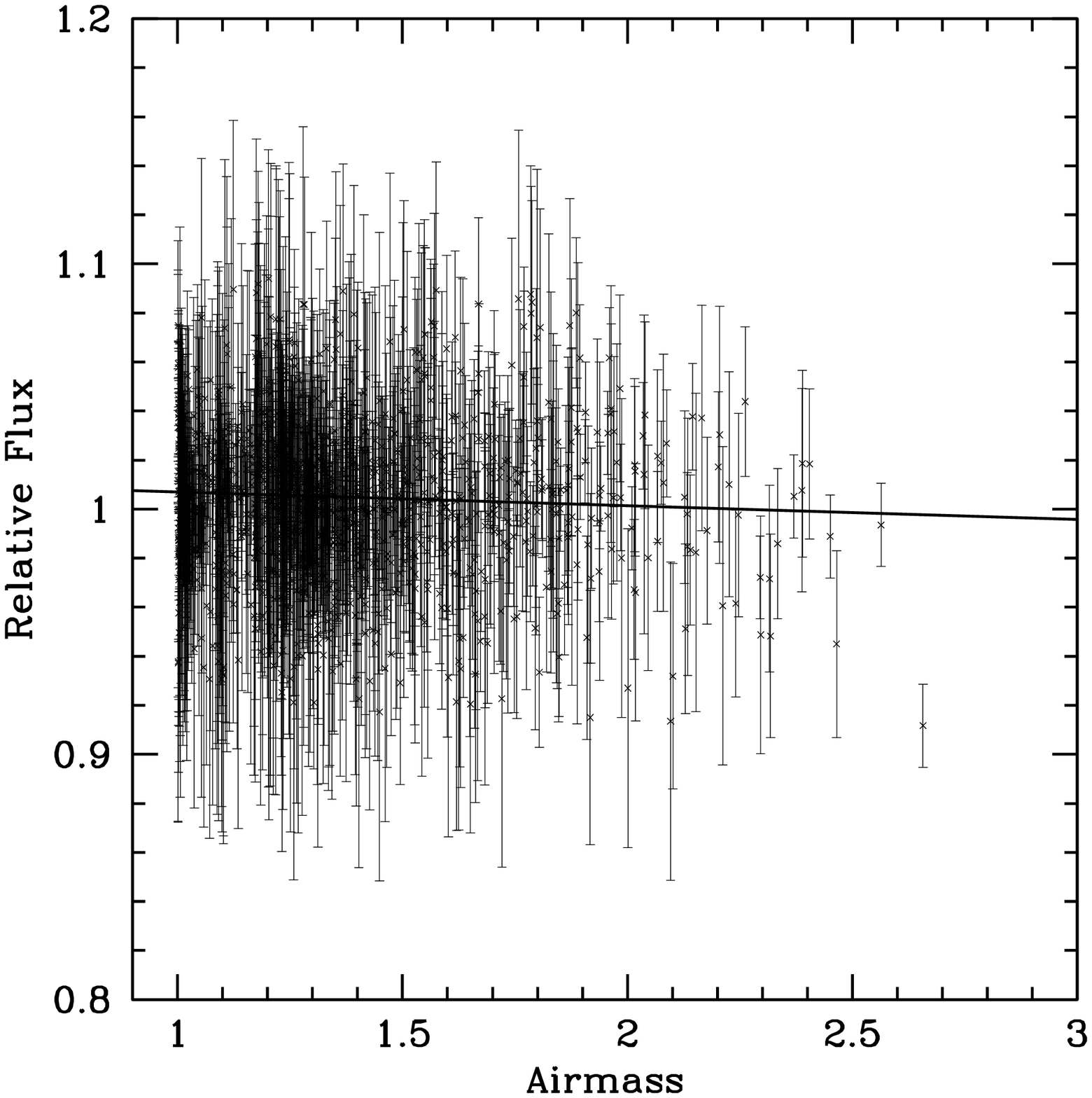}
\caption{\textbf{Left:} Transmission functions and object
spectra used in our calculation of the effect of telluric water vapor
variations on our photometry.  The bottom panel demonstrates that the
flux of a typical field star is weighted toward the blue edge of the
f814w filter's transmission, while the opposite is true of a T dwarf. 
\textbf{Right:}  Relative photometry of our T dwarfs normalized
based on the median in the airmass range 1.3--1.5 and plotted against airmass.  
The data and fit shown here are based on the iterative rejection of
the 5\% of data points that had the largest deviation from a weighted linear least-squares fit.
The weak airmass dependence suggests that telluric wavelength-dependent extinction has no
significant effect on our photometry.
\label{fig:airmass}}
\end{figure}

We model the effects of telluric PWV on our relative photometry by predicting the final
fluxes received by the CCD at airmasses ranging from 1.0 to 2.0 and telluric PWV values
ranging from 2.3mm to 10.0mm.  Although water vapor in the atmospheres of T dwarfs
themselves suppresses their emitted flux from 900-960nm, we calculate that the fraction
of their detected flux that resides in this band is still higher than for the
much bluer reference stars: thus, an increase in PWV should cause a spurious dimming
in the relative photometry of our targets.  We obtain models of atmospheric transmission from
Gemini Observatory\footnote{http://www.gemini.edu/sciops/telescopes-and-sites/observing-condition-constraints/ir-transmission-spectra; see also \citet{Lord1992}} (adopting the ones for Cerro Pachon since its
elevation is closer to Kitt Peak's than that of Mauna Kea).  The Kitt Peak website provides
both the transmission curve for our filter\footnote{http://www.noao.edu/kpno/filters/4indata/1560.html}
and the quantum efficiency (QE) curve for a typical Kitt Peak CCD\footnote{see http://www.noao.edu/kpno/manuals/dim/\#CCDchar}.
For the spectra of our field stars, we use models from Robert Kurucz' website\footnote{http://kurucz.harvard.edu/stars.html;
for one of many papers on the generation of such spectra see \citet{Kurucz2011}}.  After
confirming very similar results for stellar spectral types from A0 to K0, we adopt a K0V spectrum
to represent our average field star.  For our brown dwarf spectra we use LRIS data from
\citet{discovery}, \citet{Kirkpatrick2000}, and \citet{Strauss1999} for objects ranging in
spectral type from L7 to T6 (the T6 spectrum is actually SDSS 1624, one of our survey targets).  
The left panel of Figure \ref{fig:airmass} illustrates our calculation for this object.
Our results are necessarily approximate.  Sources of uncertainty include the long-wavelength
cutoff of the QE curve for the particular CCD we used (its exact curve was not available) and
the difference between Kitt Peak and Cerro Pachon.  We seek only a representative estimate of
the effects of changing PWV.  

Our model predicts that the measured relative flux of SDSS 1624, compared to a K0 reference star, will
change by 4.5\% at airmass 1.5 if the PWV goes from 2.3mm to 10.0mm.  The corresponding
change is 3.7\% for the late L dwarf spectra.  Even granting a large uncertainty in our
results, these values rule out PWV effects as a conceivable source of the $\sim$ 20\%
variability observed in 2M 1324.  Thus far, however, it would appear that the elevated nightly means
of brown dwarfs on February 20 and June 15 could be explained by variations in the PWV.
The change would be in the form of unusually low PWV (e.g. 2mm) on those nights, relative
to average values that would have to be higher for the other nights of the respective runs:
about 7mm for the February run and 10-15mm for June.

Airmass fits, however, allow us to also rule out this hypothesis.  Our model predicts
that in going from airmass 1.0 to 2.0, the relative flux of a T dwarf will drop
by 4.5\% if the telluric PWV is 10.0mm, and 1.8\% even if it is only 2.3mm (the
nonlinearity is likely due to saturation of some lines).  For the February run,
excluding the possibly deviant night of February 20, we find that on average
brown dwarfs actually brighten by 1.1$\pm$0.6\% over the same airmass range,
and for the June run we find they get only 0.5$\pm$0.7\% fainter.  These measurements deviate
from the predictions of our model, at respective PWV values of 7mm and 10mm, by
more than 5$\sigma$ in each case.

We conclude that PWV variations cannot explain our June 15 photometry and are
unlikely to explain our February 20 results.  Furthermore, the lack of significant airmass
dependence suggest that our model has overestimated the strength of PWV effects in
our data. A possible explanation is that the true QE curve of our detector cuts off
at a shorter wavelength than the curve we employed: our calculation is sensitive to this
parameter, and we were unable to obtain a QE curve for the specific CCD used in our observations.
Note well that f814w observations with a different CCD might be more sensitive
to PWV than those we have reported herein.

\subsubsection{Telluric Dust Absorption}
The elevated mean fluxes of brown dwarfs measured on February 20 and June 15
could in principle be caused by elevated dust content in the air above
Kitt Peak on these two nights.  This would preferentially dim the reference
stars due to their bluer colors, and thus make the brown dwarfs appear
anomalously bright.  If this has in fact happened, the measured absolute
fluxes of the reference stars should be suppressed on these nights by
a percentage considerably larger than the anomalous brightening observed
in the brown dwarfs, since the former will be an absolute effect and
the latter a differential effect measured across a relatively small range
in wavelength.  We test this hypothesis by measuring a consistent
set of bright field stars across all nights in the respective observing
runs.  

We identify a set of bright stars in the field of each 
respective brown dwarf, measure the fluxes of these stars using large apertures
of 20.0 pixel radius to avoid aperture losses, and find the average flux over the
bright stars on each image.  We find the median of these average flux measurements over a set of
images taken at low airmass in minimally clouded conditions each night. Finally,
we probe nightly changes in the transparency of Earth's atmosphere by tabulating
how much the median low-airmass fluxes of these bright stars change from one
night to the next, relative to their averages over the respective observing
runs.  We present these results in Table \ref{tab:nodust}.

\begin{deluxetable}{ccccc}
\tablecaption{Nightly Differences in Median Raw Fluxes from Bright Stars \label{tab:nodust}}
\tablehead{\colhead{Date}  & \colhead{WISE 0819 field} & \colhead{2M 0937 field} & \colhead{SDSS 1624 field} & \colhead{2M 1629 field}}
\startdata
February 18 & +2.8\% & +4.0\% &\nodata & \nodata \\
February 19 & \nodata & +0.3\% &\nodata & \nodata \\
February 20 & +0.3\% & +0.2\% &\nodata & \nodata \\
February 21 & -3.1\% & -4.5\% &\nodata & \nodata \\
June 15 & \nodata & -\nodata & +5.2\% & +5.4\% \\
June 16 & \nodata & -\nodata & -1.6\% & -1.2\% \\
June 17 & \nodata & -\nodata & -3.8\% & -3.7\% \\
June 18 & \nodata & -\nodata & +0.2\% & -0.5\% \\
\enddata
\end{deluxetable}

We find that the mean raw flux of bright stars at low airmass can vary by up to 5\% from
night to night, with typically more extinction on nights where we reported clouds 
in Table \ref{tab:data}, even though the images used for our bright-star measurements 
were substantially unclouded.  However, the extinction was not especially high on either February 20
or June 15, and the observed nightly offsets in relative photometry of our T dwarfs
do not correlate with variations in the telluric extinction.

\subsubsection{Offsets in Nightly Means are not Telluric}

We conclude that telluric effects on our relative photometry of T dwarfs are very
small.  They cannot provide a satisfactory explanation of the fact that both brown
dwarfs that we observed on February 20, and both those that we observed on June 15,
had elevated mean brightnesses on those nights.  We proceed to calculate the odds
of this happening by chance.

Excluding the June 14 observations of WISE 1516, we have 31 data sets.
They were obtained over 22 nights: two data sets each on
nine of the nights and a single data set each on the remaining thirteen nights. 
Six of these data sets had deviant means.   Assuming that any data set is
equally likely to have a deviant nightly mean, the probability that four of
these six deviant means would occur on just two
of the nights when two objects were observed is 1.68\%.  This does not take into
account the fact that on each of these nights, both objects were deviant in the same direction.
That consideration halves the probability, yielding 0.84\%.  The coincidence is therefore
unlikely, but not extremely so.  Having ruled out plausible physical causes,
attributing the coincidence to chance alignment is not unreasonable.

Thus we conclude that the nightly mean offsets of WISE 0819 and 2M 0937 on February 20,
and those of SDSS 1624 and 2M 1629 on June 15, are due to random error
or true astrophysical variability, not to telluric systematics.  For 2M 1629, the offset
is so significant that a random explanation is very unlikely, particularly
given the other evidence (see Section \ref{sec:indvar}) that this object is variable.  
Thus we maintain our classification
of 2M 1629 as a true astrophysical variable, and conclude that our observation of it
as unusually bright on June 15 is also astrophysically real.

\subsection{Probing Correlations between Seeing and Relative Photometry} \label{sec:seeing}

Photometric measurements in general can show a dependence on seeing for at least two reasons:
unresolved binarity of the target object \citep{Koen2013a}, and significant variation in the
instrumental PSF across the field.  We do not expect these effects to be significant for our observations:
none of our targets is a known binary, and the optical simplicity and small field-of-view of the 2.1m
CCD imager should render the instrumental PSF essentially constant across the field. Nevertheless
we briefly investigate correlations between our relative photometry and the seeing. We consider
primarily correlations within individual data sets (observations of one target on one night),
so that differing photometric apertures cannot mask any effects from seeing. Correlations
can be effectively explored in individual data sets because these individual sets span a considerable
range in seeing: the median RMS variation of measured seeing values within a data set is
0.21 arcsec, and the minimum and maximum RMS variations are 0.07 arcsec (2M 1629, June 16)
and 0.43 arcsec (WISE 1506, May 19). We probe for correlations within each of our 32 data
sets by performing a linear fit to relative target flux as a function of seeing
FWHM, with 3$\sigma$ clipping.

\subsubsection{Correlation results for Non-Variable Targets} \label{sec:seeconst}

We consider first the possible presence of correlations between seeing
and flux in data sets for our ten T dwarfs that are not confirmed variables. 
Among the twenty-six data sets in this category, the slope we fit
for flux versus seeing has greater than 3-sigma significance in four cases: WISE 0819 
observed on February 21 (+3.3$\sigma$), 2M 0937 on February 21
(+3.5$\sigma$), 2M 1106 on March 08 (+5.4$\sigma$), and SDSS 1254 on March 12 (+3.3$\sigma$).
Binarity of our target objects would bias our relative aperture photometry toward higher
fluxes when the seeing is poor, so these positive correlations look at first glance like
evidence of binarity.  However, WISE 0819, 2M 0937, and 2M 1106 were observed on other nights with much smaller
and/or negative correlations, contrary to what we would expect from a true unresolved binary. 
SDSS 1254 is the only target for which the positive correlation with seeing exists
in all observations: on both March 12 and (at lower significance) March 09. However, archival Hubble Space
Telescope (HST) images at 0.9$\mu$m (PropID 9844, PI: Bertrand Goldman)
rule out a companion at any separation beyond about 0.09 arcsec, which would be too small
to affect our photometry.
Since the seeing tends to change slowly on a timescale of a few hours, low-significance
astrophysical variations at typical brown dwarf rotation periods could easily produce
a spurious correlation with seeing.

Nevertheless, the slopes of our fits to relative flux versus seeing
are sufficiently strongly biased toward positive values as to suggest a physical cause. 
The weighted mean slope over the twenty-six data sets not corresponding to confirmed
variables is 1.4$\pm$0.3\%/arcsec. Even if the 3$\sigma$ significant slopes mentioned above
for WISE 0819, 2M 0937, 2M 1106, and SDSS 1254 are removed, the aggregate slope
drops only to to 0.6$\pm$0.3\%/arcsec, remaining positive and 2$\sigma$ significant.
Neither binarity nor variation in the instrumental PSF (which, even if it
exists, would not affect all targets in the same direction) appear
a viable explanation of these significant and positive aggregate slopes.
 
One other possibility exists that would create a seeing dependence having the observed
sign for relative photometry of T dwarfs in general.  Within the passband of the f814w filter, the flux of
T dwarfs is strongly weighted toward the red, while that of reference
stars is in general weighted toward the blue (Figure \ref{fig:airmass}).  
Since the size of the seeing disk is a weak inverse function of the observing
wavelength ($\mathrm{FWHM} \propto \lambda^{-1/5}$; Fried 1965),
it follows that our reference stars will have slightly larger
seeing widths than our science targets.  When the seeing worsens, the reference stars
will be affected more, causing a spurious brightening of the T dwarfs.  We suspect
that this underlying effect is in fact responsible for the statistically improbable
preponderance of positive slopes we have observed. However, full modeling to test this hypothesis is beyond the
scope of the current work.  The aggregate dependence of relative photometry on seeing
found in this analysis is too small to have a meaningful effect on our variability
analyses, so we have not corrected our photometry for it.

\subsubsection{Correlation results for Confirmed Variables}

\begin{figure}
\plottwo{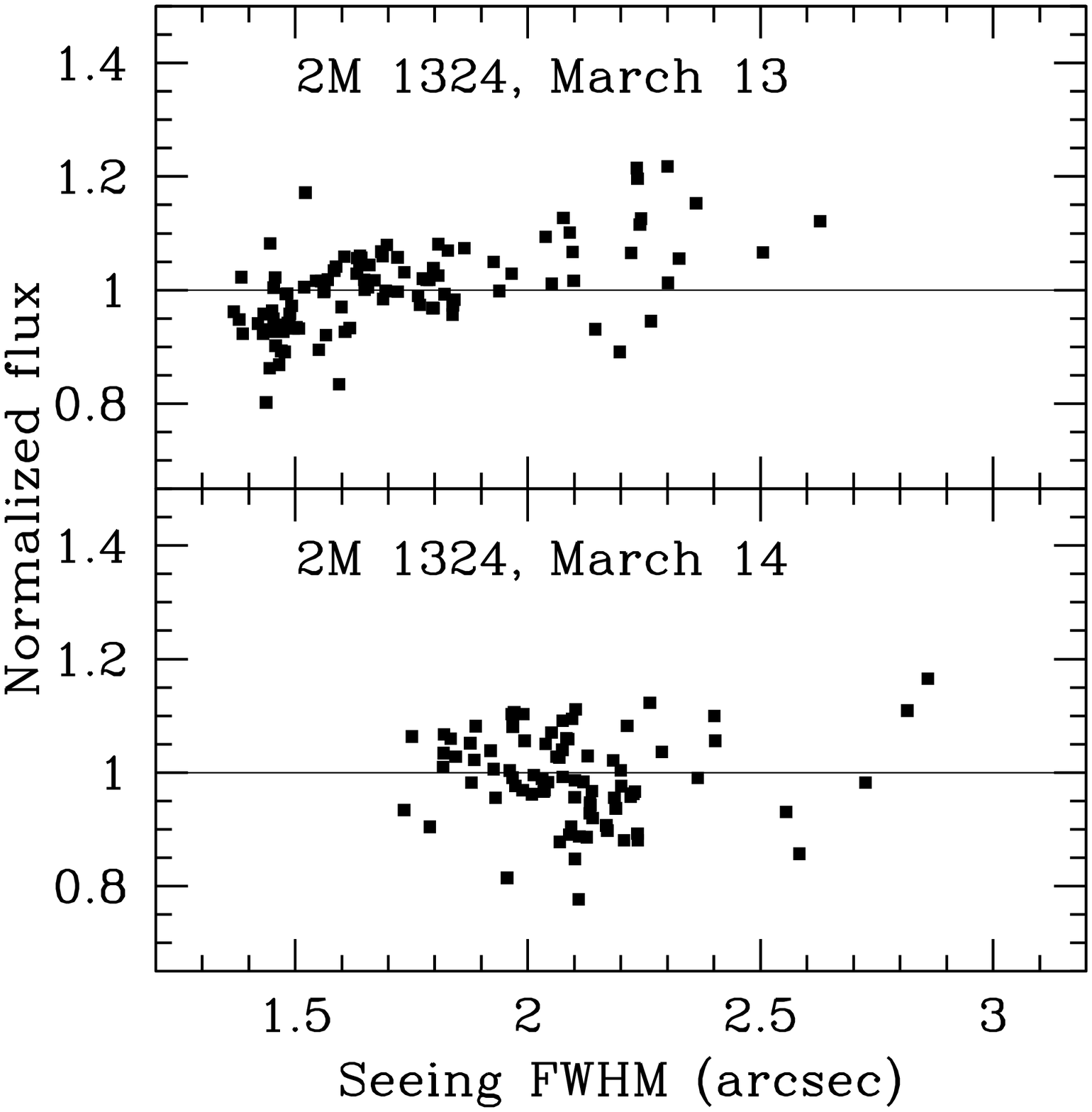}{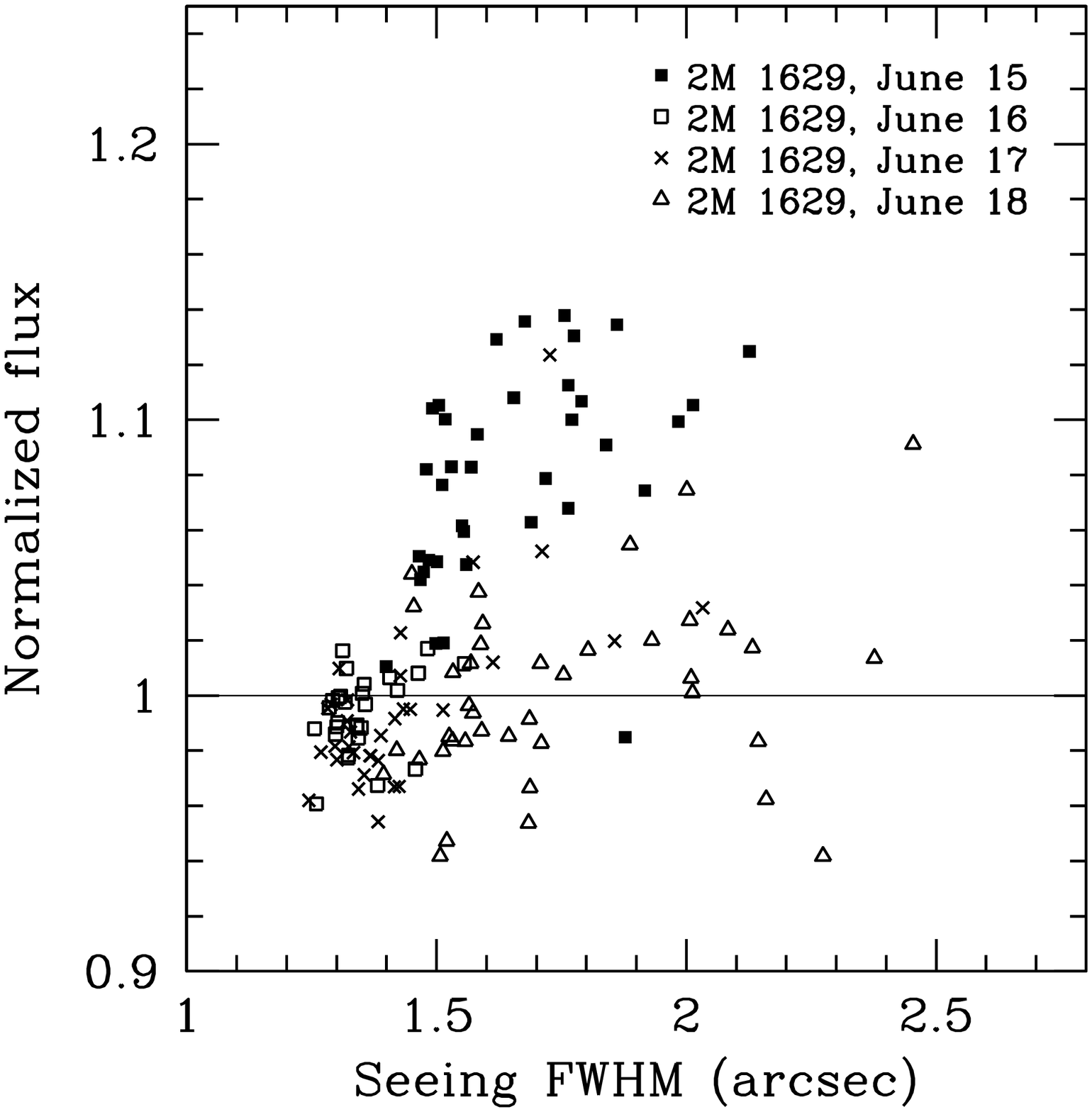}
\caption{Relative photometry plotted against seeing for the variable T dwarfs 2M 1324 (left panel) and 2M 1629 (right panel). Correlations between true astrophysical variability and seeing might be expected on individual nights, given potentially similar time scales for both phenomena.  However, they are not consistent from night to night. 2M 1324 shows a positive correlation between flux and seeing only on March 13, not on March 14. For 2M 1629, strong positive correlations are seen individually on June 15, 16, and 17; the correlation on June 18 is much less significant. However, when the photometry from all four nights is analyzed with a uniform aperture and mutually normalized (right panel), it appears as a double-valued function of the seeing: an indication that the seeing correlations on individual nights are not physically meaningful.}
\label{fig:seeing}
\end{figure}

As mentioned above, seeing tends to vary with a similar timescale to
that expected for genuine astrophysical variability of T dwarfs. We might therefore
expect the variables 2M 1324 and 2M 1629 to show correlations of photometry with seeing even if there is
no causal relationship. Significant correlations are indeed observed for
2M 1324 on one of the two nights when it was observed (March 13; 8.3$\sigma$), and
for 2M 1629 on three of the four nights (June 15--17; 3.9--4.9$\sigma$).
As would be expected if the objects were binaries, all of these correlations are positive
and are too strong to be entirely explained by the physical effect suggested in Section \ref{sec:seeconst}.
However, the lack of similarly strong positive correlations with seeing on the
other two nights of observations, March 14 for 2M 1324 (0.02$\sigma$) and
June 18 for 2M 1629 (2.3$\sigma$) does not appear consistent with a binary
hypothesis (see Figure \ref{fig:seeing}).  Indeed, archival HST images of both objects at 1.3$\mu$m 
(PropID 12550, PI: Daniel Apai; PropID 13299, PI: Jacqueline Radigan) reveal
them to be single to within 0.13 arcsec. 

To further test the existence of a meaningful correlation between brightness and seeing
specific to 2M 1629, we re-analyzed its photometry with a constant aperture across all four nights. 
If brightness and seeing were
correlated, we would expect a strong monotonic trend. Conversely, if the brightness versus
seeing correlations from individual nights are merely a coincidence of the similar time 
scales between genuine photometric variability and atmospheric changes, the correlation 
should diminish.  We plot the mutually normalized photometry 
for the four June 15--19 nights as a function of seeing in Figure \ref{fig:seeing}b.  We find that the 
overall correspondence between normalized flux and seeing is much noisier.  In fact,
the mean trend appears double-valued as a function of seeing.  Therefore, we conclude that the observed 
correlations between photometric flux and seeing on individual nights are coincidental.

\subsection{Variability of 2M 1324} \label{sec:1324}

\begin{figure}
\includegraphics[scale=0.8]{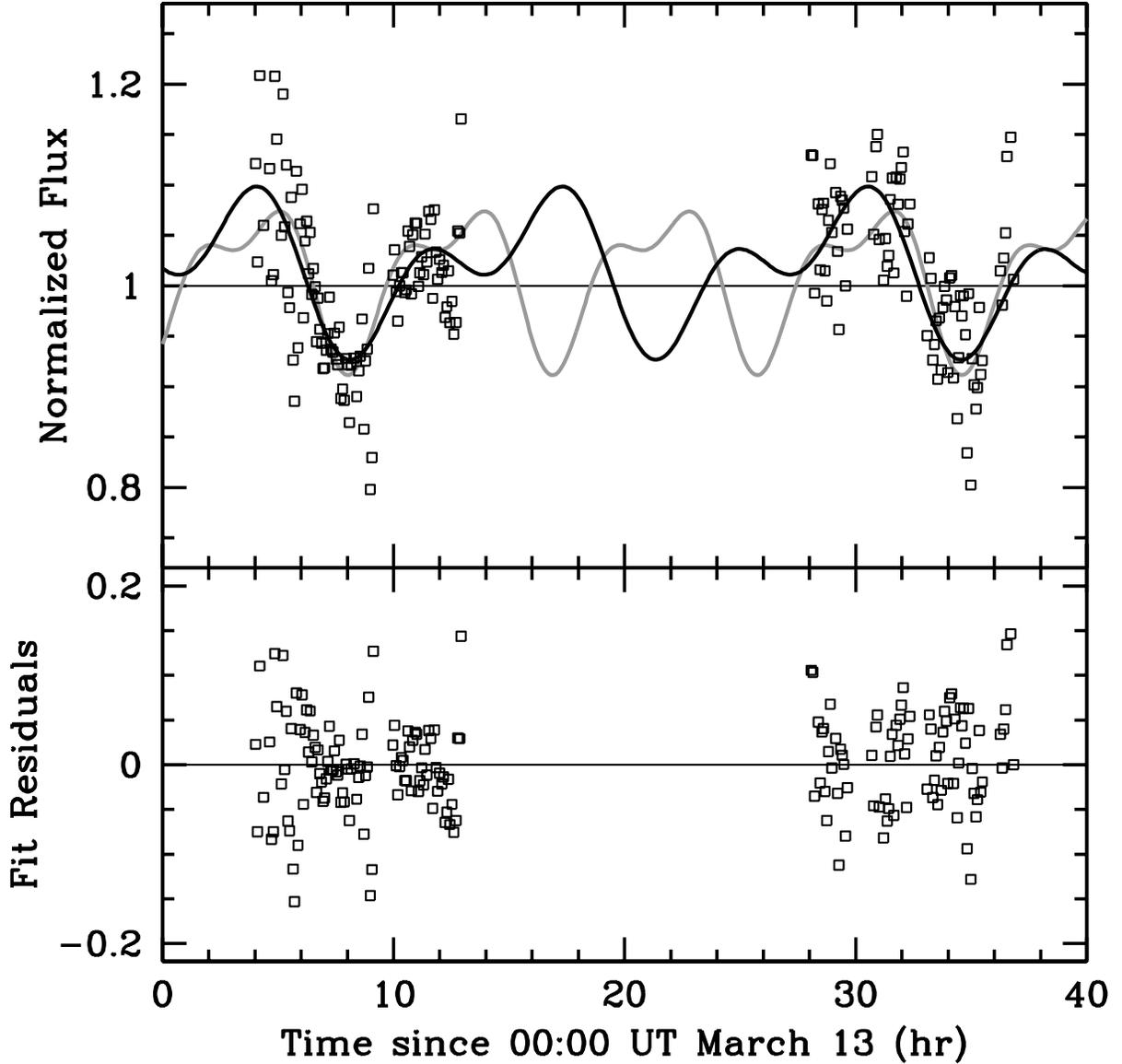}
\caption{Our full two-night photometric time series for 2M 1324,
showing two different two-term Fourier models and
residual errors.  The gray curve has a period of 8.86 hr
and an amplitude of 16.3\%, while the heavier black curve has a period of 13.23 hr
and an amplitude of 17.2\%.  Our f814w data are $2\sigma$ consistent with a range of
periods within $\pm 0.2$ hr of either value. Spitzer observations indicate
that the 13 hr solution represents the true period \citep{Metchev2014}, and thus the
residuals shown here are from the 13 hr fit.
\label{fig:longsine}}
\end{figure}

Having explored and ruled out relevant systematic effects in our
photometry due either to telluric extinction or changes in the seeing, we
now proceed to a detailed analysis of our confirmed variables, beginning
with 2M 1324. While \citet{Gillon2013} found the
variability amplitude of Luhman 16B to change rapidly from night to night,
our f814w photometry of 2M 1324 suggests that its variability may be more
consistent.  We use the methodology
described in Section \ref{sec:photres} to fit the photometry from both our observing nights
for this object simultaneously with a two-term Fourier series.  The results
indicate two families of fits with very similar $\chi^2$ values, representatives of which
are plotted in Figure \ref{fig:longsine}.  One family has periods near 9 hr and exhibits two minima
in the diurnal interval between our two nights of observations; the other has
periods near 13 hr and only one minimum in the interval. While our f814w photometry
cannot distinguish between these two families of fits, our previous observations with
the Spitzer Space Telescope yielded a period of $13 \pm 1$ hr \citep{Metchev2014}, 
demonstrating that the longer-period solution is correct.  
A Markov chain Monte Carlo (MCMC) analysis of our f814w photometry 
finds a period of $13.24 \pm 0.11$ hr and an amplitude of $17.2^{+1.1}_{-1.5}$\%.
Since low-level instrumental systematics not captured by our model may exist, the
quoted $1\sigma$ errors may not represent the full uncertainty on these measurements.

Even allowing for some additional systematic uncertainty, it appears that our
f814w photometry and the earlier Spitzer results of \citet{Metchev2014}
constitute complementary data sets for accurately determining the rotation period
of 2M 1324.  The f814w data cover a larger temporal baseline and thus deliver greater accuracy,
while the shorter but contiguous Spitzer IRAC time series was needed to resolve the ambiguity
between the 9 and 13 hr solutions.  The consistency between the two data sets, together with the apparently
random nature of the residuals from the f814w fit, indicates that the amplitude of 2M 1324 did not
change significantly over the two nights of our f814w observations.  While amplitude changes
might emerge with longer monitoring, the current data suggest that such changes
are slower in 2M 1324 than in Luhman 16B.

\citet{Metchev2014} found the amplitude of 2M 1324 to be $3.05 \pm 0.15$\% at 3.6 $\mu$m.
The variability we measure in the f814w filter is larger by nearly a factor of six.  
This difference could represent intrinsic dependence of the amplitude
on wavelength, which would likely mean that the cloud features causing the
variability have much higher contrast at $\sim$0.9 $\mu$m than at longer wavelengths. 
Amplitude differences of this type are expected theoretically, have been previously
observed in several brown dwarfs, and are useful for constraining the temperature
and structure of brown dwarf clouds \citep{SIMP0136,2M2139,Apai2013,Heinze2013,Biller2013}.
However, since our f814w measurements were not simultaneous with the Spitzer photometry, 
the amplitude difference could in principle also be explained by changes in the amplitude over time.
In any case, we find 2M 1324 to be a high-amplitude variable, comparable to Luhman 16B
and 2MASS J21392676+0220226.
It is only the third brown dwarf confirmed to show periodic variability with greater
than 10\% amplitude at any wavelength.  

\citet{Burgasser2010} identified 2M 1324 as a strong candidate unresolved binary based on its peculiar
near-infrared spectrum, which could be better matched by the blended flux of an L8 and
a T3.5 dwarf than by any single template.  As discussed in Section \ref{sec:seeing}, HST imaging of 2M 1324
shows no evidence of binarity, so if it is in fact binary, the separation must be
less than 0.13 arcsec.
Nothing in our observations of 2M 1324 is inconsistent
with its being an unresolved close binary, and in this case the component dominating
the variability would likely have the highest amplitude of any known brown dwarf.  However, we
note that an object with high-amplitude rotationally modulated variability must have a
strongly inhomogeneous photosphere, which may by itself produce a spectrum that is difficult
to match using single templates from less variable objects. \citet{2M2139} made a similar suggestion
concerning 2MASS J21392676+0220226, which was also
identified by \citet{Burgasser2010} as a candidate blended-spectrum binary, but for which 
high-resolution HST imaging did not find any detectable companion.  It is therefore possible
that high-amplitude variables characteristically show peculiar, blend-like spectra.

\subsection{Variability of 2M 1629} \label{sec:1629}

\begin{figure}
\includegraphics[scale=0.8]{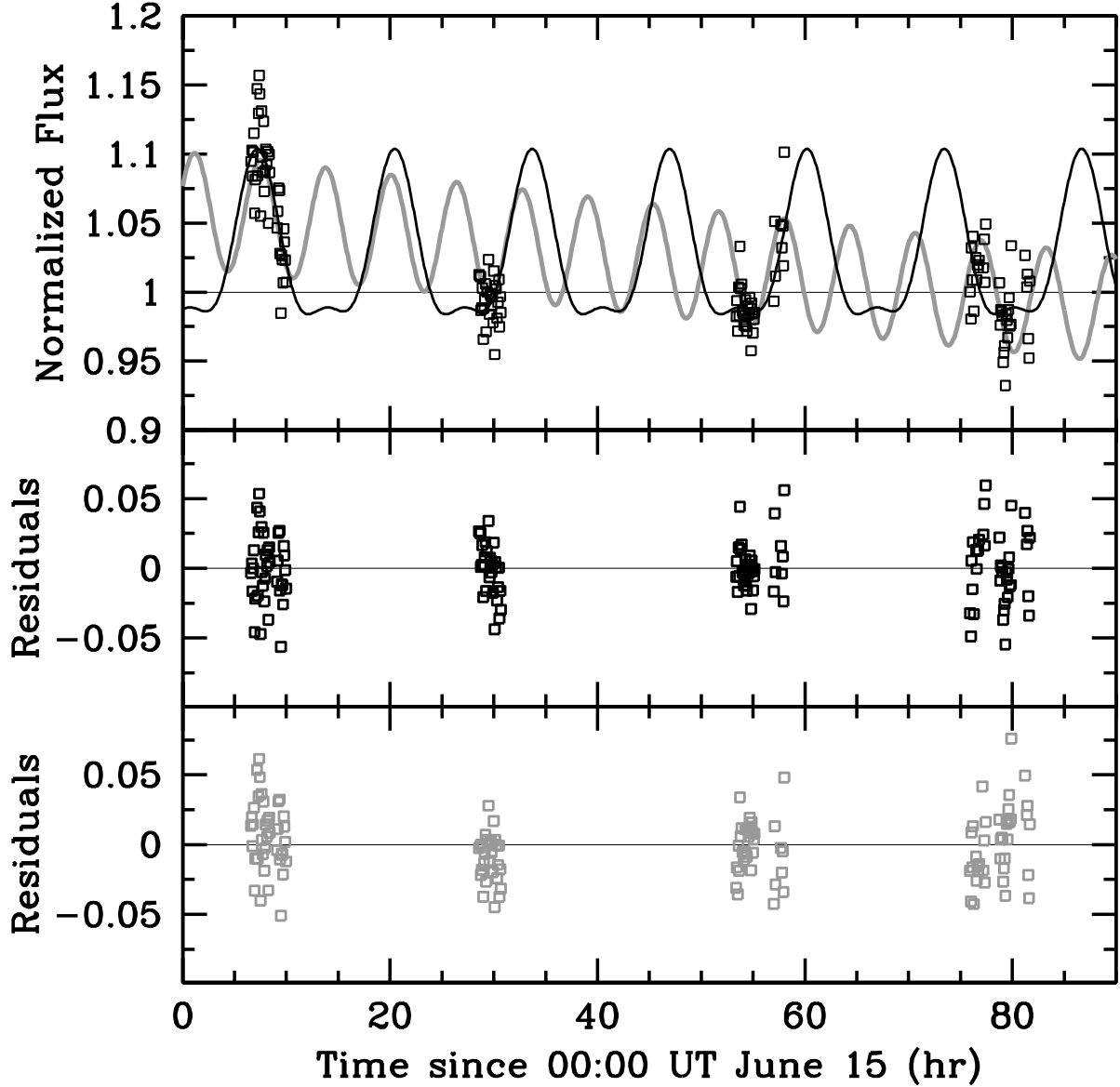}
\caption{Our full four-night photometric time series for 2M 1629,
showing two different model fits and the residuals from each.  
The black curve is a two-term Fourier fit with a period of 13 hr,
while the gray curve is a pure sinusoid with a period of 6.32 hr
consistent with the Spitzer/IRAC results reported by \citet{RadiganThesis},
which yields an acceptable fit to our data only if it is accompanied
by a linear fading of the mean flux.  Residuals are plotted in black
for the 13 hr solution and gray for the shorter-period, linear fading
model.
\label{fig:longsine2}}
\end{figure}

We attempt to fit all of our photometry of 2M 1629 with
a single periodic function, just as we did with 2M 1324, understanding that this will fail if the
amplitude changes substantially from night to night like that of Luhman 16B.
Using Fourier series with either one or two terms, we find
periods near 13 hr and amplitudes of 10--12\%.  A two-term fit is 
shown in Figure \ref{fig:longsine2}. \citet{Radigan2014} presents
$J$-band photometry of this object which suggests a period of at least
5 hr and thus could be consistent with our 13 hr fit.  However, \citet{RadiganThesis}
refers to unpublished Spitzer/IRAC photometry of 2M 1629 indicating a 6 hr rotation period.
A 6 hr period is also more likely a priori than a 13 hr period: 2M 1324 notwithstanding, 
brown dwarf rotation periods are usually not longer than 12 hr \citep{Zapatero2006,Reiners2008a,Konopacky2012}.

Fourier fits with the period constrained between 5 and 7 hr
produce acceptable residuals only if we allow the mean brightness of 2M 1629 to decrease
over time, in which case a two term Fourier series is unnecessary because a
pure sinusoid produces almost identical results.  Figure \ref{fig:longsine2} presents an
example of a sinusoidal fit with a linear fading trend.  An MCMC analysis based on this solution
gives a period of $6.32 \pm 0.03$ hr,
an amplitude of $8.4 \pm 0.7$\%, and a slope for the linear fading trend of
$-0.082 \pm 0.008$\%/hr.  Despite these small formal uncertainties, we do not believe our
photometry of 2M 1629 is as constraining as for 2M 1324.  This is in part
because the short June nights limited our nightly monitoring intervals, but
more importantly, the residuals from all our fits appear somewhat non-random --- especially
on the night of June 16 when 2M 1629 showed no significant variability. This
could be explained if 2M 1629 shows rapid variations in its
amplitude, like Luhman 16B.  Our data could then be fully consistent with the
period suggested by \citet{RadiganThesis}; and the amplitude would be greater
than 10\% on June 15 and substantially less on June 16.

While we have detected highly significant variations in 2M 1629 with a full
range considerably greater than 10\%, we are not able to measure its
rotation period with any confidence.  Without a well-defined period, the
amplitude of rotational variability also cannot be accurately determined.  
Because our sinusoidal fit based on the period from \citet{RadiganThesis}
produced an amplitude slightly below 10\%, it could be argued that 2M 1629
does not qualify as a high-amplitude periodic variable.
However, since we favor an alternative interpretation in which its amplitude
would be above 10\% on some nights, we consider it to be a second
example of Luhman 16B-like high-amplitude variability in our survey --- with the
caveat that it may be a borderline case.

\citet{Radigan2014} measure a 3\% variability amplitude for 2M 1629 in the
$J$ band, based on 4 hr of data.  While a somewhat larger amplitude might
have been measured if the data had covered a full period, it would still be
smaller than our $\sim$10\% result.  Just as for 2M 1324, results for 2M 1629
therefore suggest a higher amplitude at $\sim$0.9 $\mu$m than at longer
wavelengths. The current data cannot conclusively demonstrate this, however, 
since our monitoring was not simultaneous with that of \citet{Radigan2014}.

\subsection{Limits on T dwarf Variability Amplitudes} \label{sec:limits}

We use a Monte Carlo method, combined with the variability metric $M_1$ described in Section \ref{sec:photres},
to set upper limits on the variability amplitudes of our ten non-variable brown dwarfs.
To calculate such a limit for a given photometric data
set, we construct many realizations of simulated data with exactly the same time-sampling and
RMS error as the real photometry. Each realization contains
a sinusoidal signal with random phase but a specified period and amplitude.  We find the
metric $M_1$ for each simulated data set, and compare it with the value of $M_1$ measured
for the real data.  We adjust the amplitude of the input signal until the simulated data produces a value
of $M_1$ higher than the real data in at least 95\% of 10,000 trial realizations.  Thus
we arrive at the amplitude of a sinusoidal variation which, if present in the real object,
would have had a 95\% likelihood of producing a stronger variability detection than
was actually obtained. This constitutes our 95\% confidence upper limit on the object's variability
amplitude.

We calculate such limits at periods of 2, 5, and 13 hr, corresponding to 
short, typical, and long values for the rotation periods of brown dwarfs.  
In our Monte Carlo analysis, we measure $M_1$ only at the fixed period of the input signal, since
the best fit and the strongest $M_1$ variability detection should be found at this period.  
By contrast, for the real data sets, we measure $M_1$ based on
the best-fit solution found in a search of periods ranging from 2--50 hr (for single-night data sets) or
1-100 hr (for combined data sets).  Our limits are conservative because our
Monte Carlo analysis does not use information about
the period at which the best fit, and thus the value of $M_1$, were obtained
in the real data.  For example, a significant linear trend spanning six
hours of observation in the real data will create a large value of $M_1$ at a long best-fit period.
We will then use this $M_1$ value as input to our amplitude limit code even
when we are setting the limit at a two-hour period, which would be quite
inconsistent with the smooth linear trend.  A more sophisticated approach
might therefore set considerably lower upper limits for specific periods.  However,
issues of aliasing in the real data make it difficult to verify the accuracy 
of such methods, and our less aggressive approach sets reliable limits in
a straightforward manner.  Our limits are given in Table \ref{tab:limits}.

\begin{deluxetable}{llrrr}
\tablecaption{95\% Confidence Level Upper Limits on T dwarf Variability Amplitudes \label{tab:limits}}
\tabletypesize{\footnotesize}
\tablehead{ & & \multicolumn{3}{c}{Amplitude Limits} \\
\colhead{Object} & \colhead{Date} & \colhead{$P = $ 2 hr} & \colhead{$P$ = 5 hr} & \colhead{$P = $ 13 hr}}
\startdata
WISE 0819 & 2014-02-18 & 5.0\% & 3.6\% & 4.8\% \\
WISE 0819 & 2014-02-20 & 7.2\% & 5.8\% & 7.4\% \\
WISE 0819 & 2014-02-21 & 5.0\% & 5.6\% & 7.2\% \\
WISE 0819 & all combined & 6.4\% &  6.2\% & 6.2\% \\
2M 0937 & 2014-02-18 & 5.9\% & 5.8\% & 6.6\% \\
2M 0937 & 2014-02-19 & 5.4\% & 5.6\% & 6.2\% \\
2M 0937 & 2014-02-20 & 5.2\% & 4.9\% & 5.5\% \\
2M 0937 & 2014-02-21 & 6.0\% & 6.8\% & 7.2\% \\
2M 0937 & all combined & 4.8\% & 4.8\% & 4.9\% \\
2M 1106 & 2014-03-07 & 2.2\% & 2.2\% & 3.1\% \\
2M 1106 & 2014-03-08 & 4.8\% & 7.6\% & 8.0\% \\
2M 1106 & all combined & 2.2\% & 2.4\% & 3.0\% \\
SDSS 1207 & 2014-03-08 & 1.9\% & 2.2\% & 2.2\% \\
SDSS 1207 & 2014-03-10 & 3.6\% & 5.6\% & 4.8\% \\
SDSS 1207 & all combined & 4.1\% & 5.6\% & 4.4\% \\
2M 1231 & 2014-03-10 & 7.8\% & 9.2\% & 10.4\% \\
2M 1231 & 2014-03-11 & 6.8\% & 7.2\% & 9.0\% \\
2M 1231 & all combined & 7.6\% & 7.6\% & 9.2\%  \\
SDSS 1254 & 2014-03-09 & 8.0\% & 8.2\% & 11.4\% \\
SDSS 1254 & 2014-03-12\tablenotemark{a} & 5.0\% & 6.2\% & 8.2\% \\
SDSS 1254 & all combined & 7.1\% & 7.0\% & 7.5\% \\
2M 1503 & 2014-05-17 & 9.8\% & 16.6\% & 16.2\% \\
2M 1503 & 2014-05-18 & 2.9\% & 3.1\% & 3.8\% \\
2M 1503 & all combined & 4.4\% & 4.3\% & 5.0\% \\
WISE 1506 & 2014-05-19 & 4.9\% & 7.0\% & 9.6\% \\
WISE 1506 & 2014-05-20 & 3.7\% & 6.7\% & 7.0\% \\
WISE 1506 & 2014-06-14 & 3.8\% & 5.8\% & 9.2\% \\
WISE 1506 & all combined\tablenotemark{b} & 5.5\% & 5.7\% & 9.3\% \\
SDSS 1520 & 2014-05-21 & 2.6\% & 2.7\% & 3.4\% \\
SDSS 1520 & 2014-06-14 & 3.9\% & 5.3\% & 10.0\% \\
SDSS 1520 & all combined & 6.0\% & 6.8\% & 8.6\% \\
SDSS 1624 & 2014-06-15 & 12.2\% & 16.2\% & 38.0\% \\
SDSS 1624 & 2014-06-16 & 9.2\% & 20.5\% & 68.0\% \\
SDSS 1624 & 2014-06-17\tablenotemark{a} & 8.4\% & 13.4\% & 24.0\% \\
SDSS 1624 & 2014-06-18 & 6.9\% & 6.9\% & 16.0\% \\
SDSS 1624 & all combined & 10.6\% & 11.4\% & 15.2\% \\

\enddata
\tablenotetext{a}{This date's observations show evidence for variability, but
the data have insufficient significance for us to classify the object as a confirmed
variable.}
\tablenotetext{b}{Does not include data from 2014-06-14 due to the different
set of reference stars used on that night.}
\end{deluxetable}

The variations in amplitude limit with period
that are seen in the table are mostly due to the specifics of the time-sampling, which
can allow signals in specific ranges of phase (especially for long
periods) to be very difficult to detect.  This is especially relevant
for our June observations of SDSS 1624, when the short nights made
for brief monitoring intervals and corresponding low sensitivity
at long periods.  For this reason, we have also set limits based
on the full, combined data set for each object, spanning multiple
nights.  These can be less sensitive than some individual nights,
because night-to-night variations tend to produce higher values
of $M_1$ for the combined data sets -- however, they can yield
useful limits for sustained, long-period variability.

The upper limits we have set apply strictly only to sinusoidal variations.
Variability with a markedly non-sinusoidal character could cause
either lower (e.g., for a narrowly peaked lightcurve) or higher (e.g.,
for a square wave) values of $M_1$ at the same amplitude.  The amplitude limits
for these types of variations should be correspondingly different.
Nevertheless, we note that if we apply our upper limit analysis to the
confirmed variables 2M 1324 and 2M 1629, it reliably finds upper limits that
are above the best-fit
amplitudes we have obtained for these objects in Sections \ref{sec:1324} and
\ref{sec:1629}, despite the imperfectly sinusoidal character of
their respective variations.

As shown in Table \ref{tab:limits}, we rule out 10\% variations at a 5.0 hr
period for most of our non-variable objects on most nights.  For many
objects we set limits much lower than this.  We rule out sustained 20\% amplitude
variability at any period for all objects except our two confirmed variables.

\section{Discussion and Conclusion}

We began our survey with the observation that high-amplitude variability
appeared to be very rare in brown dwarfs, and that in this context Luhman
16B was highly unusual.  However, most previous surveys had used exclusively
infrared wavelengths (e.g. Radigan et al. 2014 and Metchev et al. 2014) or
had very few T dwarfs in their sample (e.g. Koen 2013).  Thus it was
not known how common high-amplitude variability might be in T dwarfs observed
in the red optical (0.7-0.95 $\mu$m). If variability at these wavelengths
turned out to be more common than in the near-infrared, Luhman 16B would
appear less unusual.

We have intensively monitored twelve T dwarfs with photometry in an f814w filter
covering a similar wavelength range to that in which \citet{Gillon2013} found
Luhman 16B to exhibit high-amplitude variability.  We find that T dwarfs
in general are more likely than field stars to show evidence of variability
on timescales of hours to days, and in particular
two of our targets show highly significant variations with amplitude at or above
10\%.  The T2.5 dwarf 2MASS J13243553+6358281, which was previously known to
show $\sim$3\% variations at 3.6--4.5 $\mu$m wavelengths \citep{Metchev2014}, is found to exhibit
nearly 20\% variations in the f814w filter, with a consistent $\sim$13 hr period.
\citet{Burgasser2010} identified this object as a candidate blended-spectrum binary, but
we suggest that like 2MASS J21392676+0220226 \citep{2M2139}, the peculiar spectrum of
2M 1324 may be due to an inhomogeneous photosphere rather than an unresolved binary companion.
The T2 dwarf 2MASS J16291840+0335371 is our second variable.
\citet{RadiganThesis} found it to exhibit 3--5\% amplitude variations in the $J$ band.
We find it to vary with a $\sim$10\% amplitude that may change from night to night, as does
that of Luhman 16B itself.  Our data also suggest variability in the T6 dwarf 
J162414.37+002915.6, at an overall confidence level of 94.3\%. However, as we desire to maintain
a conservative approach with negligible false alarm probability, we refrain from confirming
this object as a variable. We have carefully explored and ruled out telluric causes
for the measured variability of our objects.

Two objects, or 17\% of our twelve-object sample,
exhibit high-amplitude variability in the f814w filter -- a large change from the $\sim$1\% of
brown dwarfs found to be highly variable in previous surveys at longer wavelengths.
It therefore appears that high-amplitude ($\ge 10$\%) variability in T dwarfs is more
common in the red optical than at infrared wavelengths: Luhman 16B is not as unusual as we first supposed.  

Both variable targets in our survey were previously identified as variables in infrared surveys,
but in each case the amplitudes measured in the infrared were much smaller.
This suggests that variable T dwarfs generally have higher amplitudes in the red optical than at infrared
wavelengths.  We cannot establish this definitively for any particular
object without simultaneous multi-wavelength monitoring, and Luhman 16B itself
may be an exception \citep{Biller2013} --- nevertheless, the survey results make
it highly probable that T dwarfs' red-optical amplitudes do exceed their infrared amplitudes as a rule.
Higher amplitudes in the red optical could be due to greater contrast
in photospheric cloud features at these wavelengths relative to the infrared.  If so, the
wavelength-dependent amplitudes could be profoundly informative as to the cloud
structure of highly variable T dwarfs.  Further analysis of this possibility is beyond
the scope of the current work, but it remains a promising avenue for both theoretical and observational studies of the newly-revealed
class of high-amplitude variable T dwarfs.

The two variables we report herein, together with 2MASS J21392676+0220226 \citep{2M2139}
and Luhman 16B itself, constitute a growing sample of highly variable
T dwarfs.  Such objects are excellent targets for further studies,
including simultaneous multi-wavelength observations, spectrophotometry, and spectroscopy.

\section{Acknowledgments} 
We thank Joseph Trollo for assisting with observations and Dianne Harmer
for invaluable technical support at Kitt Peak.
This research was 
supported by NASA through the Spitzer Exploration Science Program {\it 
Weather on Other Worlds} (program GO 80179) and ADAP award NNX11AB18G.
This publication makes use of the SIMBAD online database,
operated at CDS, Strasbourg, France, and the VizieR online database (see \citet{vizier}).
This publication makes use of data products from the Two Micron All Sky Survey, 
which is a joint project of the University of Massachusetts and the Infrared Processing
and Analysis Center/California Institute of Technology, funded by the National Aeronautics
and Space Administration and the National Science Foundation.
This research has benefitted from the M, L, T, and Y dwarf compendium housed at DwarfArchives.org.
We have also made extensive use of information and code from \citet{nrc}. 
We have used digitized images from the Palomar Sky Survey 
(available from \url{http://stdatu.stsci.edu/cgi-bin/dss\_form}),
 which were produced at the Space 
Telescope Science Institute under U.S. Government grant NAG W-2166. 
The images of these surveys are based on photographic data obtained 
using the Oschin Schmidt Telescope on Palomar Mountain and the UK Schmidt Telescope.

Facilities: \facility{KPNO 2.1m}

\end{document}